\documentclass[3p]{elsarticle}
\usepackage{lineno}
\usepackage{siunitx}
\usepackage{booktabs}
\usepackage{multirow}
\usepackage{subcaption}
\usepackage{xcolor}
\usepackage{hyperref}
\usepackage{graphicx}
\usepackage{amsmath,amssymb}

\journal{Nuclear Instruments and Methods in Physics Research Section A}

\begin{document}

\begin{frontmatter}

\title{Construction, commissioning, and beam test of a pilot 3D-projection opaque water-based liquid scintillator detector}

\author[bnl,wustl]{H.~Che}
\author[bnl]{M.V.~Diwan}
\author[bnl]{S.~Gokhale}
\author[bnl,bama]{P.~Kumar}
\author[bnl]{C.~Reyes}
\author[bnl]{R.~Rosero}
\author[bama]{J.J.~Wang}
\author[bnl]{G.~Yang\corref{cor1}}
\ead{gyang1@bnl.gov}
\author[bnl]{M.~Yeh}

\cortext[cor1]{Corresponding author}

\address[bnl]{Brookhaven National Laboratory, Upton, NY 11973, USA}
\address[wustl]{Washington University in St. Louis, St. Louis, MO 63130, USA}
\address[bama]{University of Alabama, Tuscaloosa, AL 35487, USA}

\begin{abstract}
We report on the design, construction, and beam test of a pilot
three-dimensional projection detector based on opaque water-based liquid
scintillator (oWbLS).  The detector consists of an
$8 \times 8 \times 16$~cm$^3$ acrylic vessel instrumented with three
orthogonal planes of Kuraray Y11 multi-clad wavelength-shifting fibers read
out by Hamamatsu multi-pixel photon counters.  The readout electronics
are based on the CITIROC front-end boards developed for the
WAGASCI and SuperFGD detectors of the T2K experiment.  The detector was
filled with oWbLS and tested with cosmic rays and proton beams of 50, 100, 250,
and 500~MeV kinetic energy at the NASA Space Radiation Laboratory at
Brookhaven National Laboratory.  We present three-dimensional event displays
of cosmic muon and proton beam candidates, and a study of transverse light
confinement via radial charge distribution measurements.  The measured
\textcolor{black}{data show tighter light confinement than
a Geant4 simulation with a 2~cm scattering length, placing the effective
scattering length well below 2~cm and confirming effective
optical confinement of scintillation light in the oWbLS medium.}
\textcolor{black}{A first measurement of the hit-level timing resolution
using 500~MeV proton beam data yields a single-channel timing
resolution of $\sigma_t \approx 0.17$--$0.28$~ns with good photostatistics.}
These results demonstrate the viability of the 3D-projection oWbLS
technology as a scalable, fully-active detector concept for next-generation
particle physics experiments.
\end{abstract}


\end{frontmatter}


\section{Introduction}
\label{sec:introduction}

Three-dimensional fine-grained scintillator detectors have emerged as a
powerful technology for neutrino interaction measurements and calorimetry.  The
Super Fine-Grained Detector (SuperFGD) of the upgraded T2K ND280 near
detector~\cite{SuperFGD} demonstrated that an array of $1 \times 1 \times
1$~cm$^3$ optically isolated scintillator cubes, each traversed by three
orthogonal wavelength-shifting (WLS) fibers, can achieve sub-centimeter
position resolution, sub-nanosecond timing, and efficient neutron
detection~\cite{NeutronSFGD,NeutronXsec,Yang_PANIC}.  These capabilities are essential
for reducing systematic uncertainties in neutrino oscillation experiments such
as T2K~\cite{T2K}, DUNE~\cite{DUNE}, and Hyper-Kamiokande~\cite{HyperK},
and have been recognized as a priority in the DOE Basic Research Needs study
on HEP detector R\&D~\cite{BRN}.

Despite these successes, the mechanical segmentation approach has practical
limitations.  Manufacturing, quality-controlling, and assembling millions of
individual cubes is a labor-intensive process spanning years.  Once cast, the
detector properties such as spatial granularity, radiation length, and
scintillation yield are permanently fixed.

Opaque liquid scintillator offers an alternative path to fine-grained
detection~\cite{LiquidO_Nature,LiquidO_review,Michigan_WLS,oWbLS_LiquidO}.
In this approach, a highly
scattering liquid medium confines scintillation photons to a small volume
around the point of energy deposition.  A lattice of WLS fibers sampling the
volume at regular intervals captures this confined light, providing spatial
information without mechanical segmentation.  The detector granularity is
determined by the fiber pitch and the optical scattering length of the liquid,
both of which are tunable parameters.  Water-based liquid scintillator
(WbLS)~\cite{WbLS_1ton,WbLS_30ton,Theia,WbLS_Steiger} is a particularly attractive
candidate for opaque detector media, as its scintillation properties,
opacity, and metal-loading capability can be adjusted through surfactant
chemistry.

We describe the design, construction, and beam test of a pilot detector that
combines the 3D-projection readout concept proven in SuperFGD with an opaque
WbLS medium.  The detector was constructed at Brookhaven National Laboratory
(BNL) and tested with cosmic rays and proton beams at the NASA Space Radiation
Laboratory (NSRL), also located at BNL.  This work serves as a proof-of-concept
for a scalable 3D-projection opaque liquid scintillator detector technology
with broad applications in neutrino physics, collider experiments, and
rare-process searches~\cite{CheYang2v3v,NeutronSFGD}.

\section{Detector concept}
\label{sec:concept}

\begin{figure}[htbp]
\centering
\includegraphics[width=0.95\textwidth]{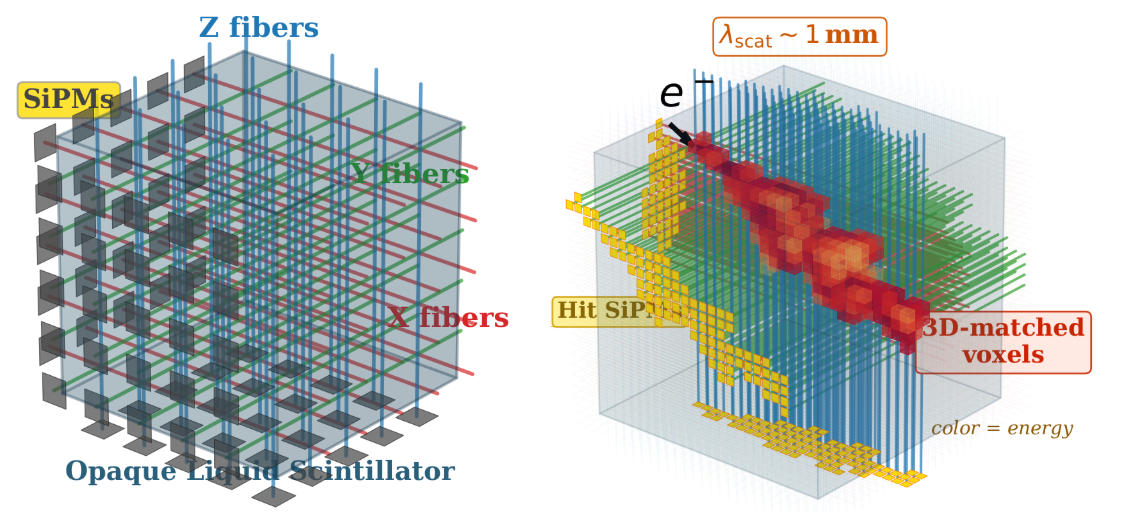}
\caption{Schematic of the 3D-projection opaque liquid scintillator detector
concept.  Left: cutaway view showing three orthogonal sets of WLS fibers
(X-fibers in red, Y-fibers in green, Z-fibers in blue) immersed in the opaque
liquid scintillator volume.  SiPMs at the fiber ends are indicated in yellow.
Right: simulated electron event showing 3D-matched voxels (red/orange cubes)
with color proportional to energy deposition, and hit SiPMs on the detector
faces (yellow).  The scattering length of the liquid is approximately
1~mm.}
\label{fig:concept}
\end{figure}

The pilot detector implements a three-dimensional projection readout in an
opaque liquid scintillator volume.  The operating principle is illustrated in
figure~\ref{fig:concept}.  Three orthogonal sets of WLS fibers, designated as
X-fibers, Y-fibers, and Z-fibers, penetrate the active volume at a regular
pitch of 1~cm.  When a charged particle traverses the liquid and deposits
energy via ionization, scintillation photons are produced isotropically.  In a
transparent medium, these photons would propagate freely throughout the
detector volume.  In an opaque medium with a scattering length of
order 1~mm, the photons undergo a stochastic random walk, forming a localized
``light ball'' around the segment of energy deposition.  The WLS fibers nearest to
the interaction point absorb the blue scintillation light (peak wavelength
$\sim$430~nm) and re-emit green photons (peak wavelength $\sim$476~nm) that are
guided by total internal reflection to multi-pixel photon counters (MPPCs) at
the fiber ends.

The three fiber views provide three independent two-dimensional projections of
the event topology: the X-fibers yield the YZ projection, the Y-fibers yield
the XZ projection, and the Z-fibers yield the XY projection.  A
three-dimensional image is reconstructed by matching hits across views that
share a common coordinate, analogous to the reconstruction technique used in
the SuperFGD~\cite{SuperFGD}.  The effective voxel size is determined by the
fiber pitch, nominally $1 \times 1 \times 1$~cm$^3$ in this pilot detector.
A detailed benchmarking study of two-view versus three-view reconstruction
for this class of detector has been reported in ref.~\cite{CheYang2v3v}.

The short scattering length of the opaque medium serves a dual purpose: it
confines scintillation light to the vicinity of the energy deposition,
providing intrinsic position sensitivity, and it suppresses optical crosstalk
between distant fibers.  This self-segmentation eliminates the need for
physical boundaries between voxels, replacing the millions of individual
scintillator cubes in a SuperFGD-type detector with a continuous,
pumpable liquid medium.

\section{Detector construction}
\label{sec:construction}

The pilot detector was designed and constructed at BNL.  The construction
proceeded in five stages: fabrication of the acrylic frame, installation of WLS
fibers, sealing of the fiber penetration points with Weld-On 40, attachment of
photosensor readout boards, and filling with opaque
liquid scintillator.  Each stage is described in detail below.

\subsection{Acrylic frame assembly}
\label{sec:acrylic}

The mechanical structure of the detector consists of a rectangular acrylic
frame defining an active volume of $8 \times 8 \times 16$~cm$^3$.  The frame
was machined from optical-grade cast acrylic (polymethyl methacrylate, PMMA)
sheets.  The acrylic boards were precision-cut and drilled with a regular
pattern of 1.0~mm-diameter holes at a pitch of 10~mm to accommodate the WLS
fibers in three orthogonal directions.  The hole pattern provides 8 fiber
positions along the X and Y dimensions and 16 positions along the Z dimension.

The acrylic boards were bonded using Weld-On solvent cement, a capillary-grade
acrylic adhesive that works through a solvent-welding process.  In this method,
the solvent softens the mating acrylic surfaces upon application; as the
solvent evaporates, the surfaces fuse into a monolithic joint.  The resulting
bond achieves tensile strengths approaching that of the bulk acrylic material.
Weld-On was selected for its optical clarity, chemical compatibility with
liquid scintillator, and proven track record in the construction of acrylic
vessels for neutrino detectors~\cite{WbLS_1ton}.  The bonding was performed in
a clean laboratory environment to avoid contamination that could degrade
optical performance.  After assembly, the joints were inspected for
uniformity, and the frame was sealed against liquid leaks.  The external edges
of the frame were reinforced with adhesive tape during assembly to maintain
alignment during the curing process.  Four threaded standoffs were installed at
the corners for mechanical mounting.
The assembled acrylic frame is shown in figure~\ref{fig:construction}~(a).

\subsection{Wavelength-shifting fiber installation}
\label{sec:fibers}

The detector is instrumented with Kuraray Y11(200) multi-clad
wavelength-shifting fibers~\cite{Kuraray} with a diameter of 1.0~mm.  The
Y11 fiber has a polystyrene core doped with K27 fluorescent dye, surrounded by
a first cladding of PMMA (refractive index $n = 1.49$) and an outer cladding
of fluorinated polymer (refractive index $n = 1.42$).  The core refractive
index is $n = 1.59$.  The multi-clad structure provides a trapping efficiency
of approximately 5\%, roughly a factor of two improvement over single-clad
fibers.  The absorption spectrum of Y11 peaks at 430~nm, well-matched to the
emission of standard organic scintillators, and the emission peaks at 476~nm
with a decay time of $7.1 \pm 0.1$~ns~\cite{KurarayFiber_new}.  The Y11 fiber
is the established standard for 3D-projection scintillator detectors, having
been used in the SuperFGD prototype beam test at
CERN~\cite{SFGD_prototype}.

The fibers were cut to lengths of 12 or 20~cm, depending on the
readout direction, allowing sufficient overhang beyond the active volume for
coupling to the photosensors.  Each fiber was individually threaded through the
pre-drilled holes in the acrylic frame.  For each of the three orthogonal
directions (X, Y, Z), the fibers were inserted in a regular grid pattern at
10~mm pitch.  The X-fibers (8 per YZ plane $\times$ 16 Z-layers = 128 fibers)
and Y-fibers (8 per XZ plane $\times$ 16 Z-layers = 128 fibers) run
horizontally, while the Z-fibers (8 $\times$ 8 = 64 fibers) run along the
beam direction.  The total number of fibers is 320.

After threading, the fiber ends protruding from the readout faces were
polished using a sequence of fine-grit sandpapers to ensure efficient optical
coupling to the photosensors.  The fiber installation is shown in
figure~\ref{fig:construction}~(b), where the dense array of fibers protruding
from the bottom of the acrylic frame is visible.

\subsection{Fiber sealing with Weld-On 40}
\label{sec:weldon40}

Weld-On 40 is a two-part reactive acrylic adhesive based on methyl
methacrylate.  After threading the fibers through the acrylic frame, the gaps
around each fiber hole must be sealed to prevent liquid leakage.
Weld-On 40 was applied around each fiber penetration point, filling the annular
gap between the 1.0~mm fiber and the drilled hole.  The adhesive cures to form
a transparent, mechanically strong, and chemically resistant seal.  Weld-On 40
was chosen over thinner solvent cements (e.g., Weld-On 3 or 4) because its
thicker viscosity prevents intrusion into the active volume while ensuring a robust
liquid-tight seal at each fiber penetration.  The curing time is approximately
15--30 minutes for initial set and 24 hours for full strength.  This sealing
step is critical: the 320 fiber penetration points represent the primary
potential leak path, in which inadequate sealing would result in oWbLS loss during
operation.

\subsection{Photosensor and readout boards}
\label{sec:sipm}

The scintillation light collected by the WLS fibers is detected by Hamamatsu
multi-pixel photon counters (MPPCs).  The MPPC model used is the
S13360-1325CS~\cite{Hamamatsu_S13360}, a ceramic surface-mount device (SMD)
featuring an active area of $1.3 \times 1.3$~mm$^2$, a pixel pitch of
25~$\mu$m, and 2,668 pixels per device.  The spectral response covers the
range 270--900~nm, well matched to the Y11 emission spectrum.  The peak
photon detection efficiency (PDE) is approximately 25\% at 450~nm; at the
Y11 peak emission wavelength and the operating overvoltage, the
effective PDE is approximately 22\%.  The breakdown voltage is approximately
$53 \pm 5$~V with a temperature coefficient of 54~mV/$^\circ$C.  The gain at the
nominal operating overvoltage is $7.0 \times 10^5$, with a typical dark count
rate of 70~kHz.

\begin{figure}[htbp]
\centering
\includegraphics[width=0.75\textwidth]{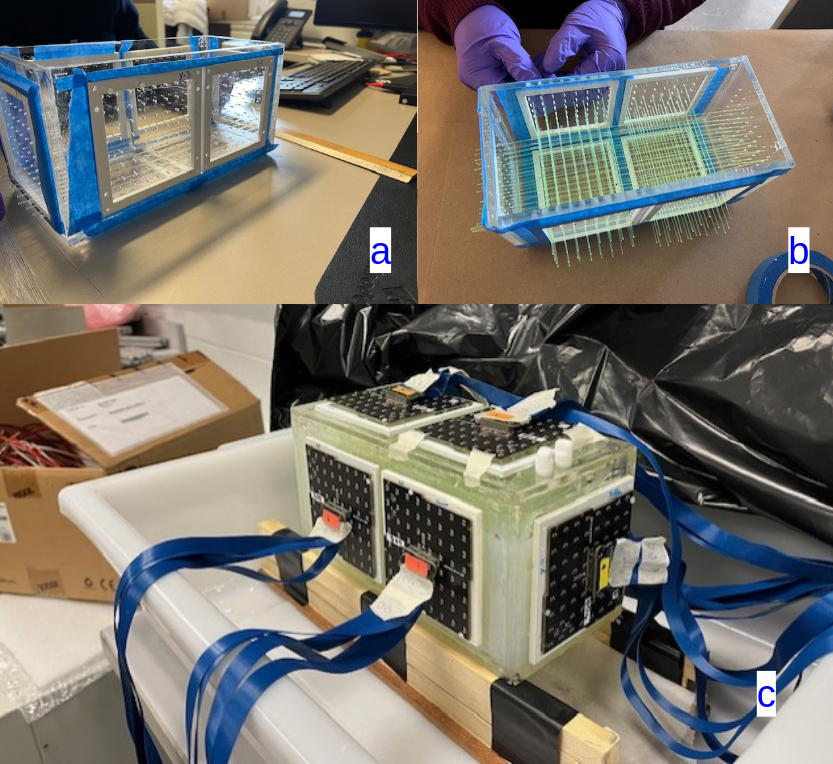}
\caption{Photographs of the pilot detector construction at BNL.
(a)~The assembled acrylic frame with internal dividers, prior to fiber
installation.
(b)~The fiber lattice structure held by hand, showing the dense array of Y11
WLS fibers protruding from the bottom of the acrylic frame after threading.
(c)~The completed detector module with MPPC readout boards ($8 \times 8$
channels each) attached to the readout faces and connected via flat ribbon
cables.}
\label{fig:construction}
\end{figure}

The MPPCs are mounted on custom printed circuit boards (PCBs) designed to
accommodate $8 \times 8$ arrays of photosensors per board, matching the fiber
grid pattern on each readout face of the detector.  The PCBs provide individual
bias voltage to each MPPC channel and route the analog signals to the
front-end electronics via flat ribbon cables.  Each readout face of the
detector is covered by one or more PCBs.  The assembled detector with the MPPC
boards and ribbon cable connections is shown in
figure~\ref{fig:construction}~(c).

\subsection{Front-end electronics}
\label{sec:electronics}

The readout electronics for this pilot detector are based on the front-end
board (FEB) system originally developed for the Baby~MIND
detector~\cite{BabyMIND_JINST,BabyMIND_construction} of the
WAGASCI experiment at T2K~\cite{WAGASCI_BabyMIND_xsec}, which
completed its first physics run in
2019--2020~\cite{BabyMIND_firstphysics}, and subsequently adapted for the
SuperFGD~\cite{SuperFGD,BabyMIND_FEB}.  This FEB system was
developed as part of the US--Japan cooperative program in high energy physics
for the T2K near detector upgrade.

Each FEB houses three CITIROC (Cherenkov Imaging Telescope Integrated Read Out
Chip) application-specific integrated circuits (ASICs)~\cite{CITIROC},
providing 96 analog input channels per board.  The CITIROC ASIC provides, for each of its
32 input channels, a charge preamplifier with configurable gain, a fast shaper
with 15~ns peaking time for timing measurements, and a slow shaper with
configurable peaking time (12.5--87.5~ns) for charge integration.  The dynamic
range spans 1 to 2,000 photoelectrons at an MPPC gain of $10^6$.  Each channel
has an individually adjustable discriminator threshold, implemented via a
10-bit common DAC with 4-bit per-channel fine adjustment, enabling triggering
down to the one-third photoelectron level.

The digitization is performed by a 12-bit, 8-channel ADC operating at
40~MS/s, which samples the analog outputs from the three CITIROC chips.  An
Altera Arria~5 FPGA on the FEB provides data formatting, trigger logic, and a
timing resolution of 2.5~ns.  The FPGA also manages the slow-control
interface for configuring the CITIROC registers and monitoring operating
conditions including temperature and humidity.

The FEBs receive signals through a fan-out distribution system.  A
synchronization subsystem distributes a common clock signal (CLK) and
synchronization pulse (SYNC) to all FEBs via dedicated fan-out boards, allowing
coherent timestamping across the entire detector.  The data acquisition is
controlled by a PC connected via a USB3 link.
The MPPC bias voltages were set to achieve the nominal overvoltage of
approximately 5~V above the individual breakdown voltage of each channel,
corresponding to operating voltages of approximately 56--58~V.

\subsection{Opaque liquid scintillator filling}
\label{sec:filling}

The liquid medium of the detector is opaque water-based liquid scintillator
(oWbLS). The manufacture of oWbLS is based on the water-based technology first developed at BNL in 2011~\cite{WbLS_1ton,WbLS_30ton,Yeh_WbLS}.  The base liquid consists of
di-isopropylnaphthalene (DIN) as the organic scintillator solvent, with
2,5-diphenyloxazole (PPO) as the primary fluor at a concentration of
5~g/L relative to the organic base\textcolor{black}{, with bis-MSB as a secondary fluor}.  Surfactant molecules encapsulate the
organic scintillator in nanometer-scale micelles, forming a stable
aqueous mixture.  In the opaque variant, the surfactant concentration and water
fraction are adjusted to deliberately maximize scattering from the
micelle structures, yielding a reduced scattering length of order a few mm at
visible wavelengths~\cite{oWbLS_LiquidO}.  The resulting liquid has a milky, translucent appearance.

The oWbLS was filled into the detector through a filling
port on the top of the acrylic frame.  The filling was performed slowly to
avoid trapping air bubbles among the fibers.  The detector was placed in a
spill tray during the filling operation as a precautionary measure.  After
filling, the detector was sealed and wrapped in opaque black plastic to
provide light-tightness.

\begin{figure}[htbp]
\centering
\includegraphics[width=0.7\textwidth]{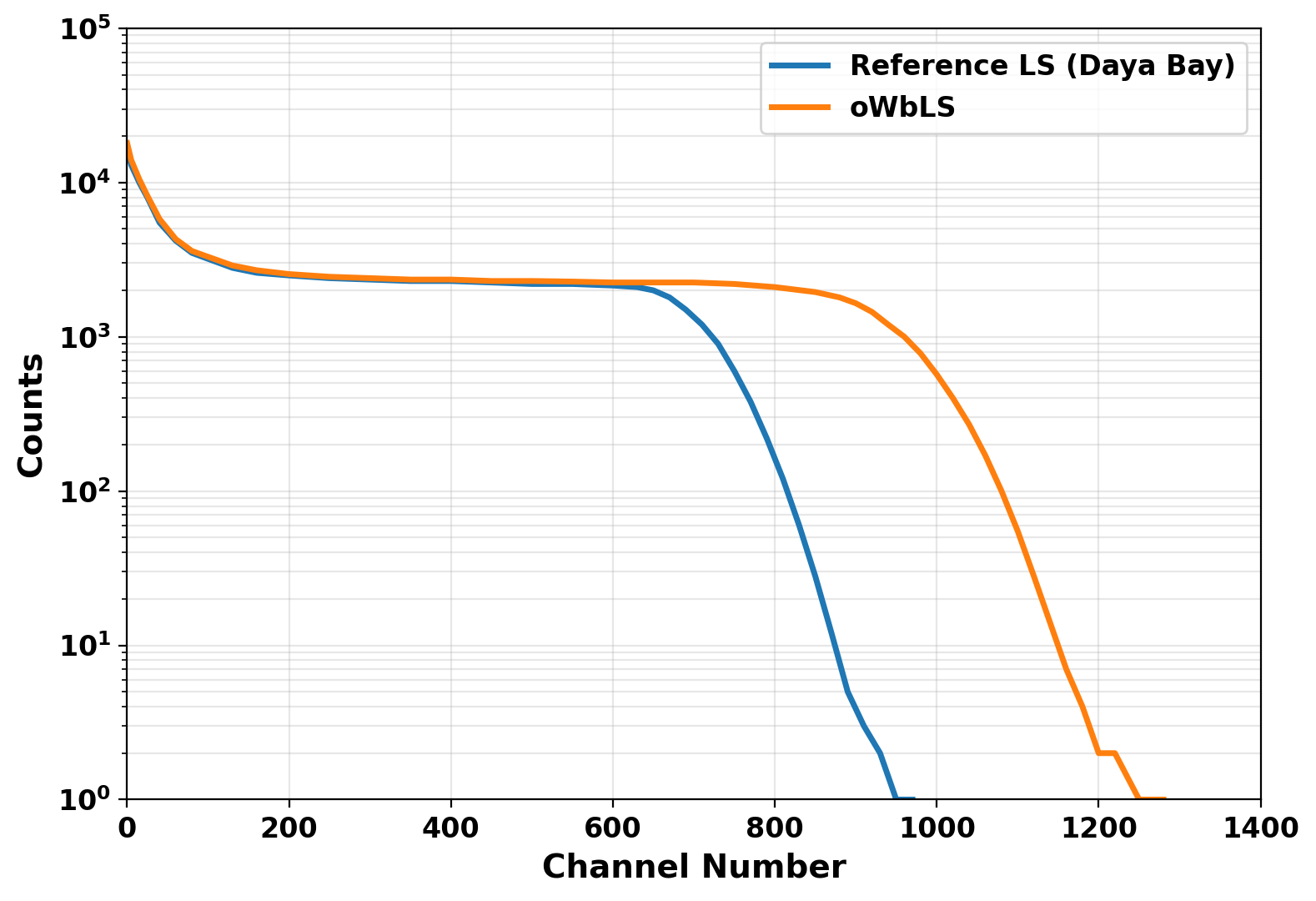}
\caption{\textcolor{black}{Compton spectra of $^{137}$Cs $\gamma$-rays
recorded with a Beckman LS6500 coincidence scintillation counter for the
reference liquid scintillator (Daya Bay LAB-based, blue) and the oWbLS
sample (orange).  The extended Compton edge position of the oWbLS sample
indicates a higher intrinsic light yield of $\sim$12,000~photons/MeV
compared to the reference value of $\sim$9,000~photons/MeV.}}
\label{fig:owbls_ly}
\end{figure}

\textcolor{black}{The light yield for the oWbLS samples was estimated relative to a
reference liquid scintillator based on well-characterized LAB systems used
in experiments such as Daya Bay~\cite{DayaBay_LS} and as summarized in
organic scintillator studies~\cite{Yeh_WbLS}.  These liquid scintillators,
composed of LAB with PPO (and bis-MSB in the case of Daya Bay), have
intrinsic light yields in the range of $\sim$8,000--10,000~photons/MeV.  In
this study, a representative value of $\sim$9,000~photons/MeV is adopted as
the reference.  The light yield measurements were carried out using a
Beckman LS6500 coincidence scintillation counter equipped with a
$^{137}$Cs gamma-ray source.  The counter employs two photomultiplier tubes
operating in coincidence mode and uses a multichannel analyzer to record
pulse height spectra.  The recorded spectra exhibit a Compton continuum
resulting from interactions of 662~keV gamma rays with the scintillator
samples.  The position of the Compton edge along the $x$-axis (channel
number) provides a qualitative measure of the scintillation light output.
By comparing the Compton edge position of an oWbLS sample to that of the
reference liquid scintillator, the relative light yield can be estimated.
The oWbLS samples studied here exhibit an estimated intrinsic light yield of
approximately $\sim$12,000~photons/MeV.  A comparison of the Compton
spectra, illustrating the relative light yield of the oWbLS samples and the
reference liquid scintillator, is shown in
figure~\ref{fig:owbls_ly}.}

\section{Beam test at NSRL}
\label{sec:beamtest}

\subsection{The NASA Space Radiation Laboratory}
\label{sec:nsrl}

The beam test was conducted at the NASA Space Radiation Laboratory
(NSRL)~\cite{NSRL}, located on the Brookhaven National Laboratory campus.
NSRL is a dedicated irradiation facility commissioned in 2003, funded by the
NASA Human Research Program, and operated by BNL.  The facility receives beams
from the BNL AGS Booster synchrotron via a 100-meter transport beam line to a
shielded target hall.  NSRL delivers proton beams with kinetic energies
continuously tunable from 50~MeV to 2.5~GeV, as well as a wide range of
heavier ion species.  The beam spill structure consists of $\sim$0.3--0.4~s
extraction spills within a 4-second cycle.  The facility operates approximately
1,000--1,200 hours per year across three run cycles.

NSRL provides a well-characterized beam with precisely known energy,
making it an ideal facility for calibrating and commissioning prototype
particle detectors.

\subsection{Experimental setup}
\label{sec:setup}

The detector was positioned on a support table aligned with the beam axis;
however, the beam entered the detector near one corner rather than at the
center, as determined by the subsequent analysis of accumulated hit maps
(section~\ref{sec:beam_profile}).  The NSRL beam is highly collimated with a
spot size well within 1~cm radius at the detector location.  The beam enters
the detector along the Z-axis (the 16~cm dimension).  The detector was wrapped
in black plastic for light-tightness.  The FEB
readout electronics and the DAQ laptop were set up in the control area adjacent
to the beam enclosure.

\begin{figure}[htbp]
\centering
\includegraphics[width=0.95\textwidth]{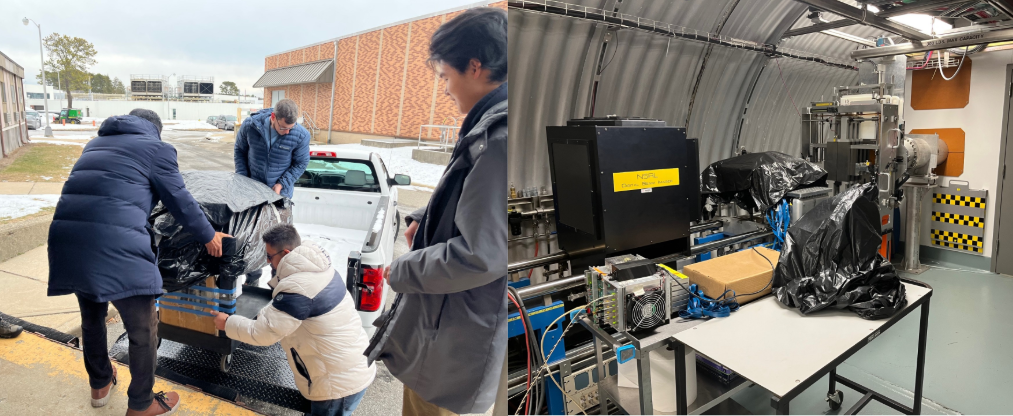}
\caption{Photographs of the beam test setup at NSRL, BNL.
Left: transport of the detector equipment to the NSRL facility.
Right: the light-tight detector (wrapped in black plastic) installed on the
beam line table in the NSRL target enclosure, surrounded by readout electronics
crates and signal cables.}
\label{fig:beamtest_setup}
\end{figure}

The detector was exposed to proton beams at four kinetic energies: 50, 100, 250,
and 500~MeV.  The 50~MeV protons have a range of approximately 2.2~cm in
water-equivalent material and stop very quickly, producing short tracks.  The
100~MeV protons have a range of approximately 7.7~cm and also stop inside the
detector.  Both the 50 and 100~MeV data sets have limited statistics and are
used primarily for event display purposes.  The 250 and 500~MeV protons are
penetrating particles in this detector: the proton range in the oWbLS medium
is approximately 37~cm at 250~MeV and 117~cm at 500~MeV~\cite{PDG_passage},
both far exceeding the 16~cm detector depth.  Each beam energy run lasted
approximately 6~minutes, with beam rates of order kHz.  In addition, cosmic
ray data were collected during beam-off periods for calibration purposes.
The transport of the detector to NSRL and the setup in the beam enclosure are
shown in figure~\ref{fig:beamtest_setup}.

\subsection{Event selection}
\label{sec:selection}

The data analysis employs a unified set of event selection criteria applied
consistently to both data and Monte Carlo simulation.  The selection cuts are:

\begin{enumerate}
\item \textbf{Photoelectron threshold:} each hit is required to exceed a
minimum of 3~photoelectrons (PE), suppressing dark-count noise and
low-amplitude crosstalk signals.
\item \textbf{Minimum hits per view:} at least 4 hits are required in each
of the two primary fiber views (XZ and YZ), ensuring that the event contains
sufficient spatial information for track or shower reconstruction.
\item \textbf{Z-layer matching:} at least 4 matched Z-layers are required
between the XZ and YZ views, where a ``match'' means that both views register
hits in the same Z-layer.  This criterion ensures consistent three-dimensional
reconstruction.
\item \textbf{Fiducial Z-range:} hits are restricted to Z-layers within the
physical detector depth, $0 \leq z < 16$~cm.
\end{enumerate}

The 3~PE threshold corresponds to approximately 3 standard deviations above the
MPPC dark noise pedestal, effectively eliminating single-photon dark counts
while retaining genuine scintillation signals.  The requirement of at least 4
hits per view ensures a minimum track length of approximately 4~cm,
corresponding to the minimum range of protons with kinetic energy above
approximately 70~MeV in water-equivalent material.  This threshold rejects
short noise clusters while accepting proton tracks at all beam energies used in
this study.  The Z-layer matching requirement of 4 layers ensures genuine
spatial coincidence between the two primary fiber views, suppressing random
combinatorial backgrounds.  These thresholds were determined empirically by
scanning their values and evaluating the signal-to-noise ratio in the proton
beam data.

\section{Data analysis framework}
\label{sec:analysis}

\subsection{Data processing pipeline}
\label{sec:pipeline}

The raw data from the FEB system are stored in binary format.  The data
processing pipeline consists of three stages:

\begin{enumerate}
\item \textbf{Unpacking:} the raw binary data are decoded into ROOT~\cite{ROOT}
tree structures containing per-channel ADC counts and timestamps.
\item \textbf{Calibration:} the ADC counts are converted to photoelectron
(PE) units using per-channel gain constants from LED calibration
(section~\ref{sec:led_calib}).
\item \textbf{Event building:} calibrated hits are organized into event
objects, each containing collections of hits indexed by fiber view and
spatial coordinate.
\end{enumerate}

\subsection{Three-dimensional reconstruction}
\label{sec:3dreco}

The three-dimensional reconstruction proceeds by matching hits between the
XZ and YZ fiber views.  For each Z-layer, the algorithm collects all hits
$(x_i, \text{PE}_i)$ from the Y-fiber (XZ) view and all hits
$(y_j, \text{PE}_j)$ from the X-fiber (YZ) view.  Every combination
$(x_i, y_j, z)$ is formed as a candidate three-dimensional voxel.  The charge
assigned to each voxel is the average of the PE values from the two
contributing views: $\text{PE}_{\text{voxel}} = (\text{PE}_{xz} +
\text{PE}_{yz})/2$.  When a Z-fiber hit is available at the corresponding
$(x_i, y_j)$ position, it provides an additional constraint that suppresses
combinatorial ghost hits, that is, false three-dimensional candidates arising from
ambiguities in the two-view projection~\cite{CheYang2v3v}.

The resulting set of three-dimensional voxels is stored in a 3D histogram
that accumulates charge per voxel position.  For track-like events, a
principal component analysis (PCA) is applied to the voxel positions to
extract the event trajectory and assess the track linearity.

It is important to note the geometry of the Z-fibers relative to the proton
beam.  The Z-fibers run parallel to the beam direction (the 16~cm dimension).
A proton entering along the Z-axis illuminates essentially one Z-fiber for the
entire track length, depositing a large integrated charge in that single
channel.  This makes the XY view (Z-fiber readout) less informative for track
reconstruction of beam protons: it registers the beam transverse position but
provides no longitudinal segmentation along the beam direction.  For this
reason, the proton beam analysis relies primarily on the XZ and YZ views
(X-fiber and Y-fiber readout), which provide the longitudinal segmentation
needed to resolve the energy deposition profile along the beam direction.  The
XY view from Z-fibers remains valuable for suppressing ghost hits in the 3D
matching and for cosmic muon reconstruction, where tracks are not aligned with
any fiber direction.

\section{Simulation}
\label{sec:simulation}

A Geant4-based~\cite{Geant4} Monte Carlo simulation framework was developed
to model the detector response.  The simulation implements the full detector
geometry, including the acrylic frame, fiber lattice, and liquid scintillator
volume.

\subsection{Detector geometry}
\label{sec:simgeo}

The active volume is modeled as an $8 \times 8 \times 16$~cm$^3$ rectangular
box filled with liquid scintillator.  The chemical composition of the
scintillator is approximated as C$_9$H$_{10}$ with a density of
1.03~g/cm$^3$.
The bulk absorption length of the liquid scintillator is set to 10~m.

\textcolor{black}{The Y11 WLS fibers (section~\ref{sec:fibers}) are modeled with
their multi-clad optical structure.  The fiber pitch is 10~mm in all three
directions, matching the physical detector.  The bulk absorption length in the
fibers is set to 4.0~m, and the absorption length in the acrylic frame is
5.0~m.}

The scintillation emission spectrum peaks at \textcolor{black}{$\sim$430~nm}, with a configurable
scintillation yield (nominally \textcolor{black}{12,000}~photons/MeV) and a time constant of
\textcolor{black}{1.3~ns}.  The WLS fiber absorption is strongly wavelength-dependent, ranging
from 0.5~mm at 354~nm to 10~m at 620~nm, with re-emission peaking at 540~nm
(green) and a WLS time constant of 7.0~ns.

The MPPC photosensors are modeled as 0.5~mm-diameter sensitive discs at the
fiber readout ends, with a photon detection efficiency (PDE) of 25\%.

\begin{figure}[htbp]
\centering
\includegraphics[width=0.95\textwidth]{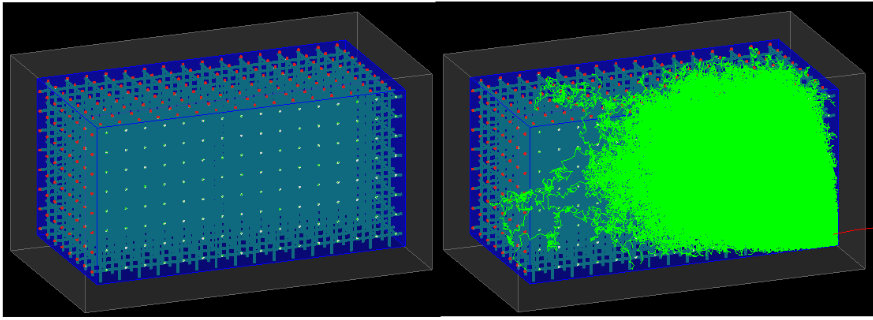}
\caption{Geant4 visualization of the pilot detector simulation.  Left: the
full detector geometry showing the acrylic vessel (gray, semi-transparent),
the liquid scintillator volume (blue), three orthogonal fiber arrays (green),
and MPPC photosensors (red) attached to the left readout panel.  Right:
optical photon propagation from a 50~MeV proton entering from the right corner
with a scattering length of 1~cm.  The green tracks show the trajectories of
individual scintillation photons undergoing multiple scattering events before
being absorbed by a WLS fiber or escaping the volume.}

\label{fig:sim_vis}
\end{figure}

The Geant4 visualization of the simulated detector is shown in
figure~\ref{fig:sim_vis}.

\subsection{Optical physics}
\label{sec:simoptics}

The simulation tracks individual optical photons from production to detection.
Scintillation photons are generated isotropically at each energy deposition step.
Scattering in the liquid is implemented with a tunable scattering
length parameter.  This parameter controls the mean free path between elastic
scattering events, which randomize the photon propagation direction.
Simulations were generated with scattering lengths of 0.4, 0.5, 0.6, 0.8,
1.0, 2.0, 5.0, 10.0, 100.0, and 1000.0~mm to bracket the expected value for
the oWbLS medium.

Absorption in the bulk scintillator, wavelength shifting in the Y11 fibers
(absorption at 430~nm, re-emission at 476~nm), and total internal reflection
in the fiber cladding are all modeled.  The MPPC photon detection efficiency
in the simulation is set to 25\%, matching the measured PDE of the
S13360-1325CS at the Y11 emission wavelength and the operating overvoltage
used in this experiment.

\subsection{Digitization and analysis}
\label{sec:simdigitize}

The simulated optical photon hits at the MPPC surfaces are digitized using the
same algorithm applied to data.  The digitized simulation output is passed
through the identical event selection and three-dimensional reconstruction
pipeline described in sections~\ref{sec:selection} and~\ref{sec:3dreco},
ensuring a consistent comparison between data and Monte Carlo.

\section{Calibration and light yield}
\label{sec:cosmic}

\subsection{LED gain calibration}
\label{sec:led_calib}

The MPPC gain calibration was performed using a pulsed LED system in late 2020,
approximately five years prior to the beam test campaign at NSRL.  The LED
produces low-intensity light pulses that illuminate each MPPC channel,
generating a charge distribution with well-resolved single-photoelectron (SPE)
peaks corresponding to 0, 1, 2, 3, \ldots\ detected photons.  The separation
between consecutive peaks in the ADC charge spectrum provides a direct
measurement of the single-PE gain $G$ (in ADC counts per photoelectron) for
each channel, independent of the LED pulse intensity.  This SPE peak-spacing
method is intrinsically robust: it measures the MPPC gain
$G = C_{\text{pixel}} \times \Delta V / e$, which depends only on the pixel
junction capacitance (a fixed geometric property of the device) and the
applied overvoltage.

The calibration was performed for all MPPC channels at the nominal operating
bias voltage and at a controlled temperature of approximately 22$^\circ$C.
The resulting per-channel gain constants are used throughout this analysis to
convert raw ADC counts to photoelectron units.

\subsection{MPPC gain stability over five years}
\label{sec:gain_stability}

The LED calibration was conducted approximately five years before the beam test
data taking.  We assess the potential systematic effect of this time gap on the
PE scale.

The single-PE gain of a SiPM is determined by
$G = C_{\text{pixel}} \times (V_{\text{bias}} - V_{\text{BR}}) / e$, where
$C_{\text{pixel}}$ is the pixel junction capacitance (a geometric property of
the silicon, fixed at fabrication), $V_{\text{bias}}$ is the externally applied
bias voltage, and $V_{\text{BR}}$ is the breakdown voltage.  Any intrinsic gain
drift over time must therefore originate from a shift in $V_{\text{BR}}$, as
$C_{\text{pixel}}$ is invariant and $V_{\text{bias}}$ is set by the readout
electronics.

Published studies of multi-year SiPM operation provide strong evidence that
intrinsic $V_{\text{BR}}$ drift is negligible in the absence of radiation
damage.  The MAGIC telescope collaboration operated SiPM camera modules
continuously for eight years (2015--2023) and reported no long-term
deterioration trend in SiPM gain after correcting for temperature
correlations~\cite{MAGIC_SiPM_aging}.  The T2K experiment operated
approximately 60,000 Hamamatsu MPPCs in the ND280 near detector since 2009,
with a total failure rate of only 0.5\% over the full experiment lifetime; all
observed light yield reduction was attributed to plastic scintillator and WLS
fiber aging, not to MPPC degradation~\cite{T2K_scint_aging}.  The JUNO-TAO
experiment performed accelerated burn-in tests on over 4,000 Hamamatsu SiPM
tiles, confirming stable breakdown voltage
characteristics~\cite{JUNO_TAO_SiPM}.

On these grounds, the intrinsic MPPC gain drift over five years of
room-temperature storage is expected to be well below 1\%.

The dominant source of gain variation between the 2020 calibration and the
2026 beam test is temperature.  The Hamamatsu S13360-1325CS has a breakdown
voltage temperature coefficient of 54~mV/$^\circ$C.  At the nominal operating
overvoltage of 5~V, a temperature difference $\delta T$ between calibration
and beam test conditions produces a fractional gain shift of
$\Delta G / G \approx 54 \times \delta T / 5000 \approx 1.1\%$ per $^\circ$C.
If the laboratory temperature during the beam test differed from the
calibration temperature by $\pm$2$^\circ$C, a conservative estimate for an
indoor facility, the gain would shift by approximately $\pm$2.2\%.

If the effective gain at beam test is lower than at calibration (e.g., due to
a higher ambient temperature increasing $V_{\text{BR}}$), the 2020 calibration
constants would slightly overestimate the single-PE gain, causing the reported
PE values to underestimate the true photoelectron yield.  Conversely, a lower
temperature at beam test would cause a slight overestimate of PE.  We estimate
the resulting systematic uncertainty on the absolute PE scale at approximately
2--3\%, dominated by the unmonitored temperature difference.

This systematic does not affect the transverse radial charge profiles presented
in section~\ref{sec:radial}, which are normalized to the total integrated
charge and are therefore insensitive to an overall PE scale factor.  Nor does
it affect the best-fit scattering length determination, which depends on the
shape of the radial profile rather than its absolute normalization.  The light
yield measurement of 13--14~PE per MIP (section~\ref{sec:lightyield}) carries
this $\pm$2--3\% systematic uncertainty, which is small compared to the
$\sim$30\% channel-to-channel variation observed across the detector.

We note that both the 2020 LED calibration and the 2026 beam test were
conducted in temperature-controlled laboratory and accelerator environments
at BNL, where the ambient temperature is maintained at approximately
22--25$^\circ$C.  The actual temperature difference between the two
measurements is therefore expected to be well within the $\pm$2$^\circ$C range
assumed above, and the corresponding systematic is likely closer to 1\% than
to the conservative 2--3\% upper bound.

\subsection{Light yield from cosmic muons}
\label{sec:lightyield}

The light yield of the oWbLS medium is characterized using cosmic muon data.
A cosmic muon traversing the detector vertically passes through approximately
1~cm of liquid scintillator at each fiber plane, depositing approximately
2~MeV of energy as a minimum-ionizing particle (MIP).  For each fiber channel,
the distribution of PE values from many such muon traversals is accumulated.
The resulting distribution has a characteristic Landau-like shape, peaked at a
most probable value (MPV) with an asymmetric tail toward higher PE values,
reflecting the stochastic nature of ionization energy loss in thin absorbers.
We fit each distribution with a Landau function convolved with a Gaussian
resolution function to extract the MPV.  The measured MPV is typically
13--14~PE per fiber per MIP traversal, corresponding to approximately
13~PE/MeV per fiber for a 1~cm path length.

\begin{figure}[htbp]
\centering
\includegraphics[width=0.95\textwidth]{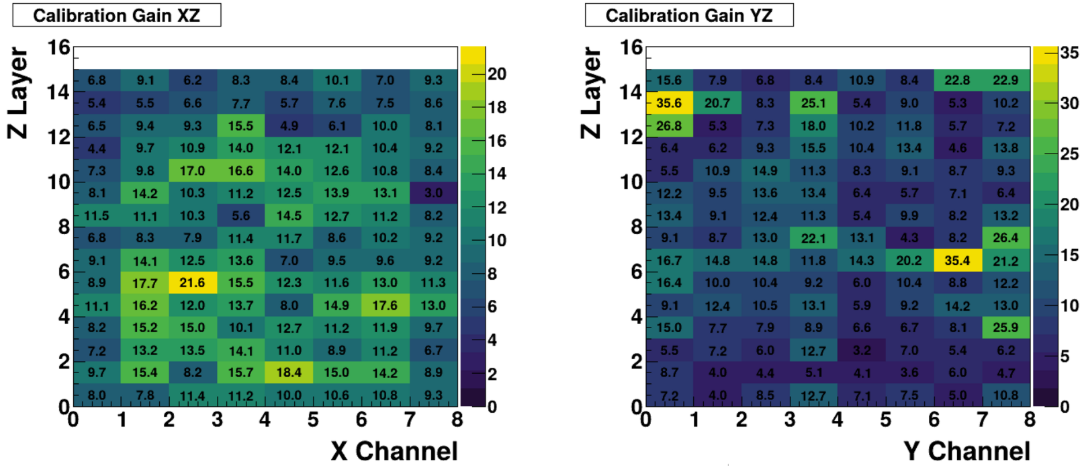}
\caption{Two-dimensional maps of the mean PE per channel for cosmic data.
Top row: XZ and YZ views from cosmic muon data.
Bottom row: XZ and YZ views from proton beam data.
The color scale represents the mean number of photoelectrons.}
\label{fig:lightyield_maps}
\end{figure}

Detailed 2D maps of the mean PE per channel for both cosmic and beam data
are provided in figure~\ref{fig:lightyield_maps}.  These maps confirm that
the light yield is reasonably uniform across the active volume, with the
highest values observed in the central region of the detector.

A small number of channels show PE values significantly above or below the
typical range; these outliers are attributed to the limited cosmic muon
statistics available in this initial data set.  The pilot detector has since
been drained to study the long-term compatibility between the oWbLS and the
fiber and acrylic materials.  A dedicated high-statistics cosmic muon
calibration campaign is planned for the next-generation prototype detector,
which will provide more precise per-channel calibration constants.

The channel-by-channel response from cosmic muon data is used to assess the
uniformity of the detector response.  Cosmic muons traversing the detector
vertically deposit approximately 2~MeV/cm as minimum-ionizing particles
(MIPs).  For each channel $(i, z)$ in the XZ and YZ fiber views, the PE
distribution from cosmic muon hits passing through that channel is accumulated.
The mean of each distribution provides the average
number of photoelectrons per MIP traversal at that position.
The gains are typically in the range
5--20~PE/MIP for the XZ view and 3--16~PE/MIP for the YZ view.  The observed
channel-to-channel variation reflects differences in MPPC sensitivity, fiber
optical quality, and coupling efficiency, with a few clear outliers resulting
from limited cosmic statistics.

\subsection{Cosmic muon event displays}
\label{sec:cosmic_display}

\begin{figure}[htbp]
\centering
\includegraphics[width=0.6\textwidth]{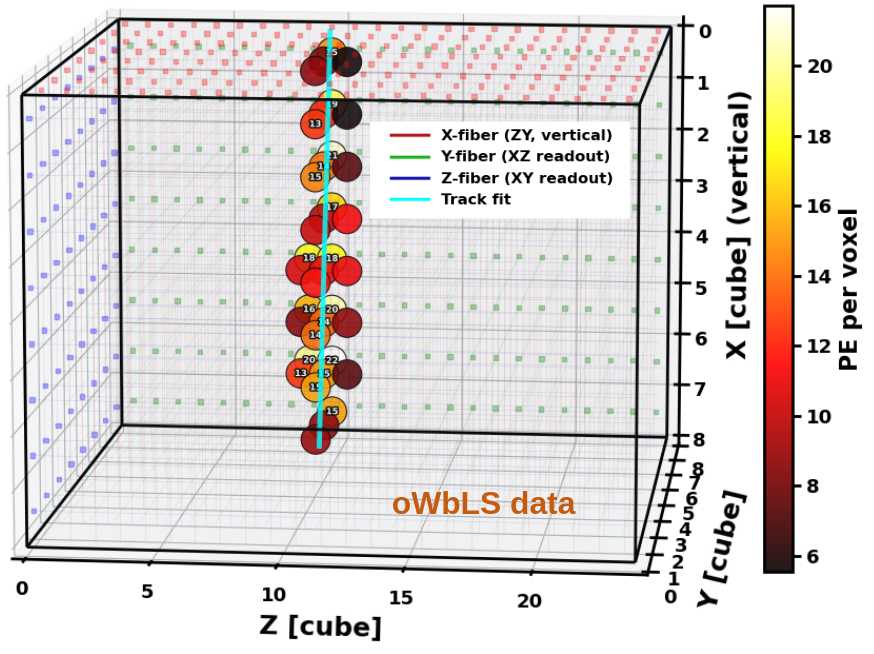}
\caption{Three-dimensional event display of a cosmic muon candidate in the
oWbLS detector.  Spheres represent 3D-matched voxels, with color proportional
to the average PE (scale on right).  The cyan line shows the fitted straight
track.  Fiber directions are indicated: X-fibers (red), Y-fibers (green),
Z-fibers (blue).  MPPC positions are shown as colored squares on the
detector faces.}
\label{fig:cosmic_event}
\end{figure}

Three-dimensional event displays of cosmic muon candidates are shown in
figure~\ref{fig:cosmic_event}.  Cosmic muons traversing the detector
vertically produce straight tracks of 3D-matched voxels.  Each voxel is
displayed as a sphere at the reconstructed $(x, y, z)$ position, with color
proportional to the average PE from the two contributing fiber views.  A
linear fit to the voxel positions is overlaid on the display, confirming the
track-like topology.  Typical cosmic muon events consist of 10--20
reconstructed 3D voxels.  The PE values for individual voxels range from
approximately 7 to 27~PE, reflecting the combined effects of Landau
energy-loss fluctuations, channel-to-channel response variations, and the
geometry of the muon track relative to the fiber grid.

\section{Proton beam results}
\label{sec:proton}

\subsection{Proton event displays}
\label{sec:proton_display}

\begin{figure}[htbp]
\centering
\includegraphics[width=0.95\textwidth]{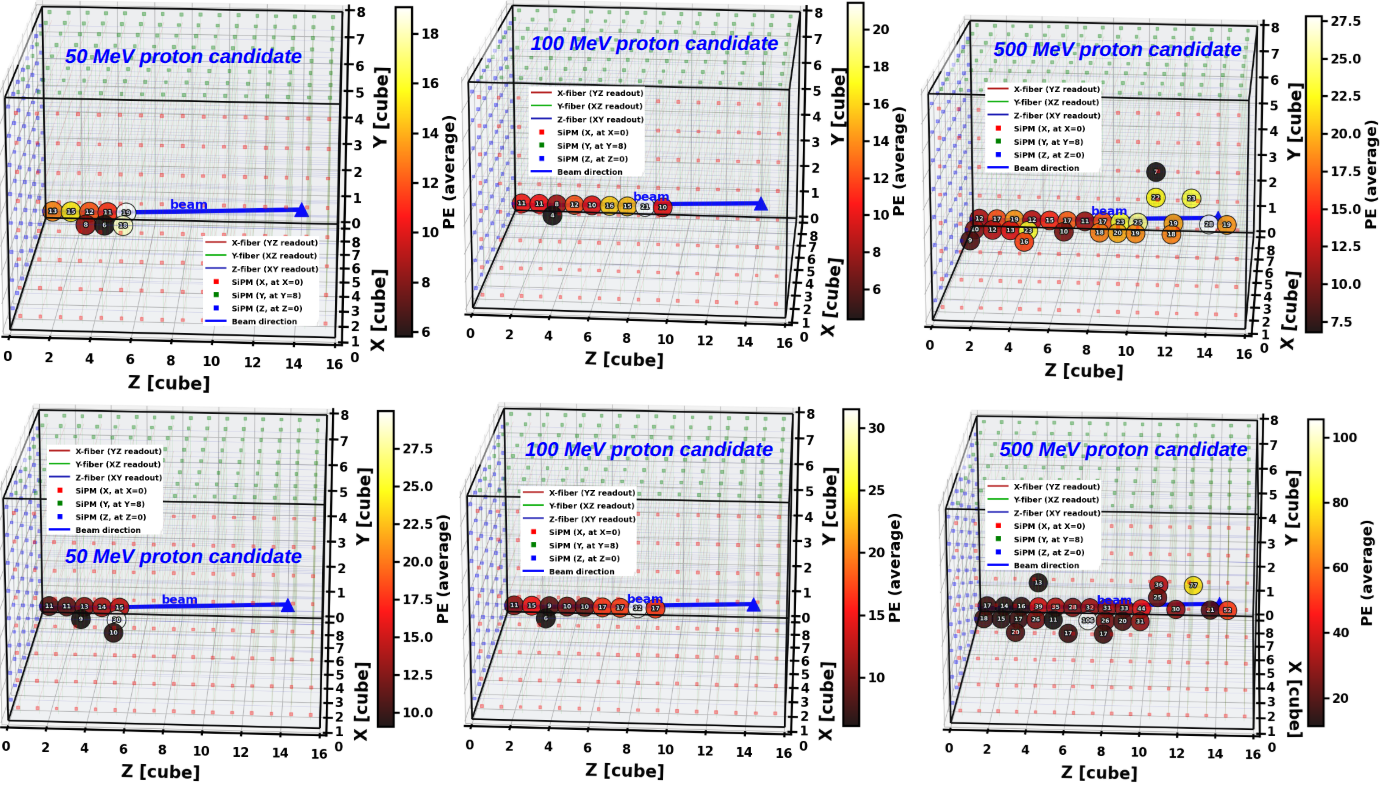} 
\caption{Three-dimensional event displays of proton beam candidates at three
kinetic energies.
Top row: 50~MeV (left), 100~MeV (center), and 500~MeV (right).
Bottom row: additional events at each energy.
Spheres represent 3D-matched voxels with color proportional to PE (scale
shown).  The beam enters along the positive Z-direction.  Statistics boxes show
the number of reconstructed voxels and mean positions.  The 50 and 100~MeV
protons stop inside the detector (ranges $\sim$2.2 and $\sim$7.7~cm,
respectively); the 500~MeV protons are fully penetrating (range $\sim$117~cm).
The 250~MeV data, which has the highest statistics and most stable beam
conditions, is shown in detail in figure~\ref{fig:proton_annotated}.
The 50 and 100~MeV data sets have limited statistics.}

\label{fig:proton_events}
\end{figure}

Three-dimensional event displays of proton candidates at 50, 100, and
500~MeV kinetic energy are shown in figure~\ref{fig:proton_events}.  The
proton tracks are oriented along the beam direction (Z-axis).  At 50~MeV,
the proton range in the oWbLS medium is approximately 2.2~cm~\cite{PDG_passage},
and the proton stops very quickly, producing a short track; the 50~MeV data
set has limited statistics.  At 100~MeV,
the proton range is approximately 7.7~cm~\cite{PDG_passage},
and the proton stops inside the detector, exhibiting a stopping-proton topology;
the 100~MeV data set also has limited statistics.
At 500~MeV, the proton is fully penetrating (range $\sim$117~cm),
producing approximately 56 reconstructed voxels with an essentially constant
energy loss rate of $\sim$3.1~MeV/cm.  Note that in the event displays, the
Z-axis extends beyond the physical detector boundary at 16~cm due to the
display coordinate convention.

\begin{figure}[htbp]
\centering
\includegraphics[width=0.65\textwidth]{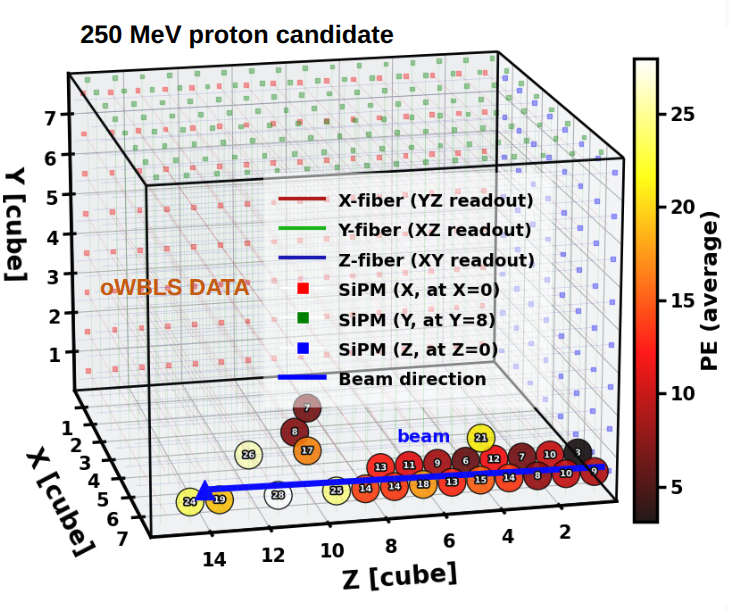}
\caption{Annotated 3D event display of a 250~MeV proton candidate in the
oWbLS detector.  Each 3D-matched voxel is displayed as a sphere with its
PE value labeled.  The blue arrow indicates the beam direction.  Fiber
directions are color-coded: X-fibers (red), Y-fibers (green), Z-fibers
(blue).  SiPM positions are shown as colored squares.  The PE variation
across the voxels reflects both the energy deposition profile and
channel-to-channel response differences.}
\label{fig:proton_annotated}
\end{figure}

The 250~MeV data, which has the highest statistics and most stable beam
conditions, is shown in detail in figure~\ref{fig:proton_annotated}.  At
250~MeV, the proton range ($\sim$37~cm) far exceeds the 16~cm detector
depth, and the proton traverses the full detector volume.

A publication-quality event display of a 250~MeV proton candidate is shown
in figure~\ref{fig:proton_annotated}, with each voxel annotated with its PE
value.  The beam direction is indicated by an arrow.  PE values range from
approximately 3 to 24 across the voxels, with the variation reflecting both
the energy deposition profile and channel-to-channel response differences.

\subsection{Accumulated beam profiles}
\label{sec:beam_profile}

The NSRL proton beam is well-collimated, with a spot size measured to be within
1~cm radius at the detector position.  The beam enters the detector near the
corner, at approximately $x \approx 7$--$8$~cm and $y \approx 1$--$2$~cm,
rather than at the geometric center.  This off-center positioning, while not by
design, provides a useful geometry: the beam track is surrounded by several
fiber channels on one side and fewer on the other, allowing the transverse
light spread to be studied over a range of perpendicular distances from the
beam axis.

\begin{figure}[htbp]
\centering
\includegraphics[width=0.95\textwidth]{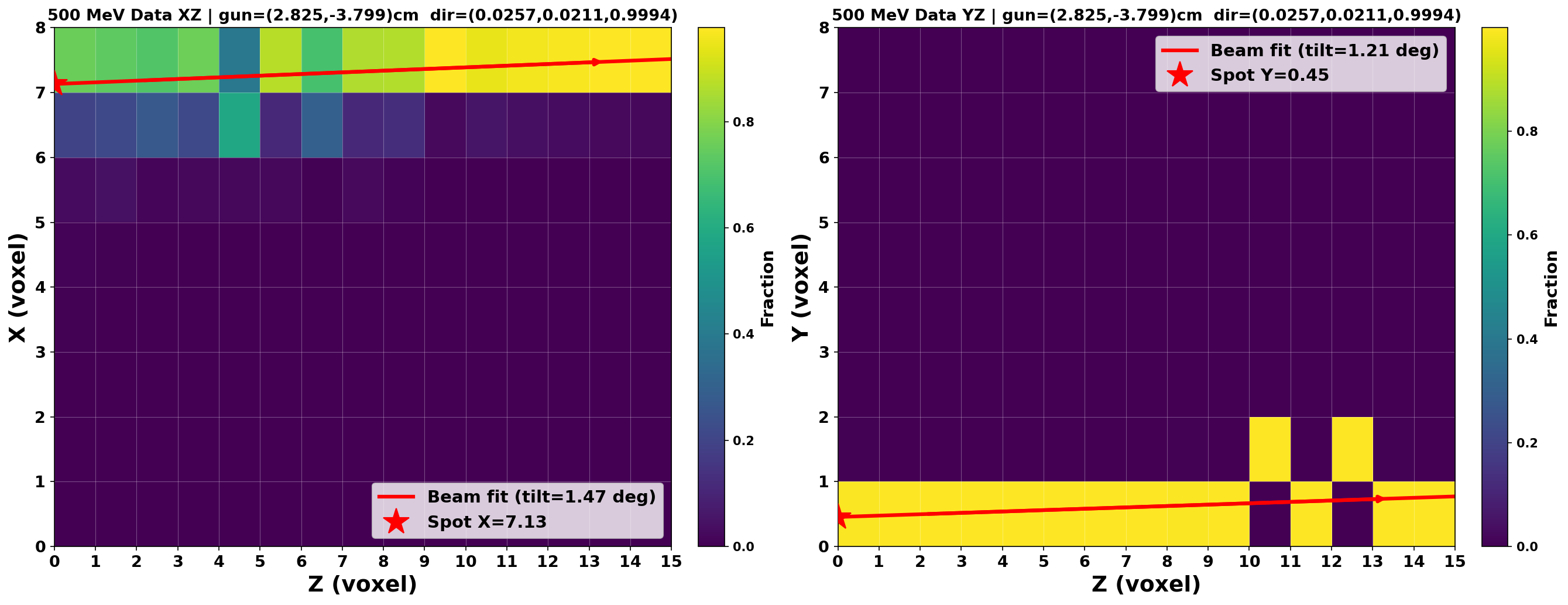}
\caption{Accumulated hit maps from 250~MeV proton beam data.
Left: XZ projection showing the beam entering at $x \approx 7$--$8$~cm.
Center: YZ projection showing the beam at $y \approx 1$--$2$~cm.
Right: 3D event display of a representative event showing the energy
deposition profile.
The color scale represents accumulated charge in units of $10^3$~PE.}
\label{fig:beam_profile}
\end{figure}

The accumulated hit maps from the 250~MeV proton beam data are shown in
figure~\ref{fig:beam_profile} for the XZ and YZ projections.  The charge is
approximately uniform along the beam direction, consistent with penetrating
protons whose energy loss rate varies only gradually across the 16~cm detector
depth.  The beam spot is well-collimated, spanning approximately 1--2 fiber
channels in the transverse directions.

A three-dimensional accumulated event display of a single representative
event is also shown, illustrating the energy deposition profile along the
beam direction.
The beam entry position was extracted by fitting the accumulated hit profiles
in three dimensions using the data from all proton events.  The fitted beam
trajectory has an entry point at $(x, y, z) = (7.32, 0.5, 1.97)$~cm with a
direction vector $(0.0148, 0.0192, 0.9997)$, indicating a slight angular
offset of $\sim$1.4$^\circ$ from the Z-axis.
\textcolor{black}{The detector and beam collimator positions were not changed between
energy settings, so the beam enters at the same location and direction for
all runs.  This allows direct comparison of transverse charge profiles across
beam energies without correcting for differences in beam geometry.}

\section{Transverse light spread studies}
\label{sec:transverse}

A central objective of this beam test is to characterize the transverse spread
of scintillation light in the oWbLS medium and to extract the effective
scattering length.  The short scattering length of the opaque
liquid is the key parameter that determines the spatial resolution and the
degree of optical confinement in this detector concept.

\subsection{Z-normalized fraction maps}
\label{sec:znorm}

To visualize the transverse spread, we compute the Z-normalized fraction
for each channel.  For a given Z-layer, the fraction is defined as the charge
in each transverse channel divided by the total charge in that Z-layer.  In a
perfectly confined medium, only the channel(s) directly traversed by the beam
would register a non-zero fraction.  Transverse light spread manifests as
non-zero fractions in channels adjacent to the beam position.

The 250~MeV proton beam was selected as the primary data set for the transverse
spread analysis for two reasons: it provides the highest statistics among the
beam energy settings, and the NSRL beam conditions were most stable during the
250~MeV run, with fewer beam interruptions compared to the other energy
settings.

\begin{figure}[htbp]
\centering
\includegraphics[width=0.95\textwidth]{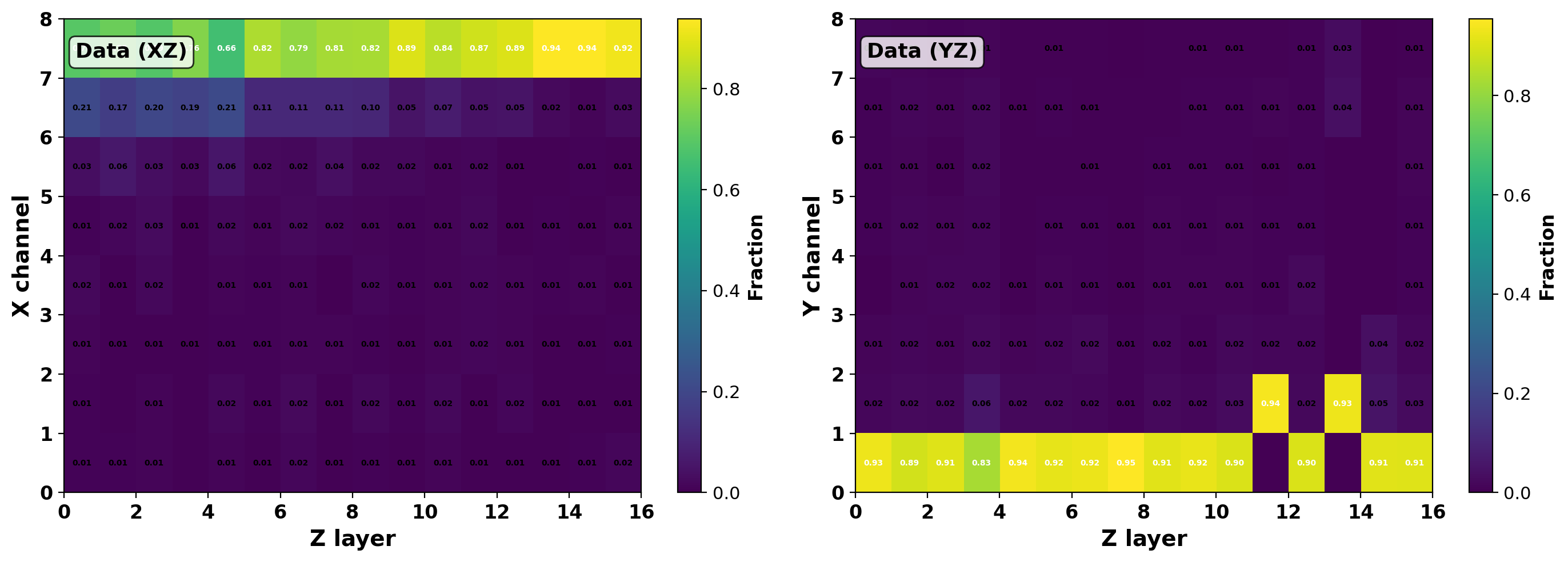}
\caption{\textcolor{black}{Z-normalized fraction maps for 250~MeV proton beam data.
Left: XZ view.  Right: YZ view.
The color scale represents the fraction of the total Z-layer charge
contained in each channel.  The data shows strong confinement, with peak fractions
of 0.94 (XZ) and 0.96 (YZ) in the beam channel, indicating that nearly all
scintillation light is captured within the traversed fiber.}}

\label{fig:znorm}
\end{figure}

The Z-normalized fraction maps for the 250~MeV proton beam data are shown in
figure~\ref{fig:znorm}.
In data, the beam track appears as a bright band at the expected transverse
position, with a Z-norm fraction of 0.7--0.9 in the beam channel and
essentially zero elsewhere.

\begin{figure}[htbp]
\centering
\includegraphics[width=0.95\textwidth]{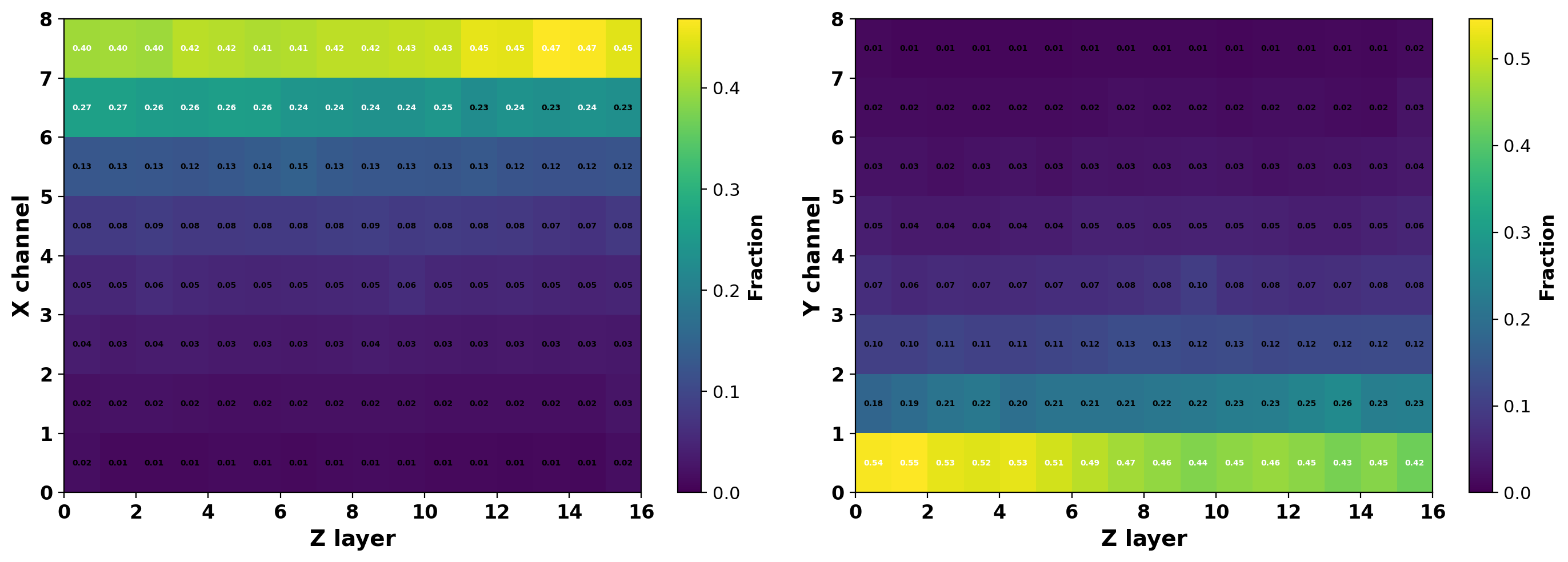}
\caption{\textcolor{black}{Monte Carlo Z-normalized fraction maps for 250~MeV proton simulation
with a scattering length of 2~cm.  Left: XZ view.  Right: YZ view.  
}}

\label{fig:znorm_mc}
\end{figure}

The Monte Carlo Z-norm fraction maps are produced by generating 250~MeV proton
events in the Geant4 simulation with a given scattering length, processing the
simulated optical photon hits through the identical digitization and
reconstruction pipeline used for data, and computing the per-layer charge
fractions in the same manner.  The Monte Carlo, generated with a scattering
length of 2.0~cm, reproduces the overall pattern but predicts a somewhat
broader distribution (Z-norm fraction of 0.35--0.50 in the beam channel),
indicating that the simulation can be further tuned.  The Monte Carlo Z-norm
fraction maps are shown in figure~\ref{fig:znorm_mc}.

\subsection{Radial profile analysis}
\label{sec:radial}

The transverse light spread is quantified more precisely through the radial
charge profile.  For each reconstructed 3D voxel, the perpendicular distance
$R$ from the fitted beam trajectory is calculated.  The charge in each voxel
is accumulated as a function of $R$, and the resulting distribution is
normalized to the total integrated charge.

\begin{figure}[htbp]
\centering
\textcolor{blue}{\includegraphics[width=0.6\textwidth]{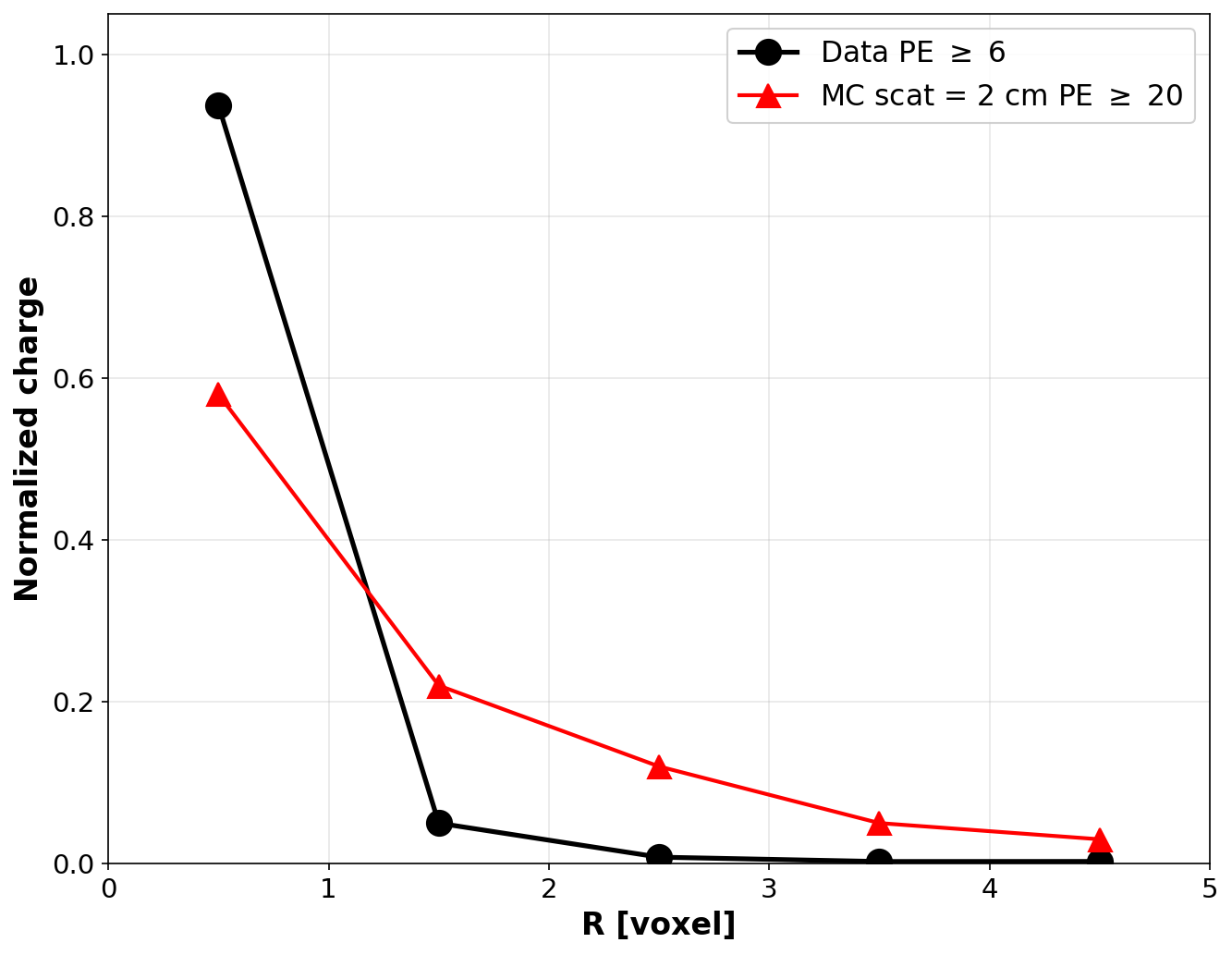}}
\caption{\textcolor{black}{Radial charge profile for 500~MeV proton beam data
(black circles) compared with a Geant4 Monte Carlo prediction with a
scattering length of 2~cm (red triangles).  The radial distance $R$ is
measured perpendicularly from the fitted beam trajectory in units of voxel
pitch (1~cm).  Both data and MC are normalized to unit integral.  Events are
selected with Z-match $\geq 4$ and data PE $\geq 6$, MC PE $\geq 20$.  The
data are more confined than the 2~cm MC, demonstrating that the effective scattering length of the oWbLS
medium is well below 2~cm.}}
\label{fig:radial_profile}
\end{figure}

\textcolor{black}{Figure~\ref{fig:radial_profile} shows the measured radial
profile for the 500~MeV proton beam data compared directly with a Geant4
Monte Carlo prediction generated with a scattering length of 2~cm.  The MC
sample is processed through the identical digitization and reconstruction
pipeline used for data, with matched Z-match ($\geq 4$). The PE thresholds were chosen to be different in data and simulation in order account for the photon collection efficiency difference in data and simulation. A wide range of MC threshold cuts have been tested and the results are very similar.}

\textcolor{black}{The data (black circles) exhibit a sharply peaked
distribution, with approximately 94\% of the normalized charge contained in
the innermost radial bin ($R < 1$~cm).  Only $\sim$5\% of the charge
appears in the first neighboring bin ($1 < R < 2$~cm), and the charge
beyond $R = 3$~cm is consistent with zero.  In contrast, the 2~cm MC
simulation predicts only less charge in the innermost bin,
with a substantially broader tail, in the second bin and
measurable charge out to $R = 5$~cm.}

\textcolor{black}{The data thus show significantly tighter optical confinement
than the 2~cm scattering length simulation.  This comparison places a
direct upper bound on the effective scattering length of the oWbLS medium:
the scattering length is well below 2~cm, consistent with the mm-scale
scattering lengths expected from the surfactant-stabilized micelle structures
in this formulation.}

Notably, both the 250 and 500~MeV data sets show the same degree of
confinement, despite a $\sim$50\% difference in their
respective $dE/dx$ values ($\sim$4.1~MeV/cm at 250~MeV versus
$\sim$3.1~MeV/cm at 500~MeV).  This consistency confirms that the observed
confinement reflects intrinsic optical properties of the oWbLS medium
rather than an artifact of the energy deposition profile.
Due to the relatively higher statistics and more stable beam conditions at
500~MeV, we adopt the 500~MeV data as the primary result.  The 250~MeV data
serves as a valuable cross-check, confirming the energy independence of the
measurement.

{\color{black}
\section{Timing resolution with 500~MeV proton beam}
\label{sec:timing}

The CITIROC-based front-end electronics provide hit-level timing information
via the fast-shaper discriminator output, digitized by the FPGA with a time
step of 2.5~ns (section~\ref{sec:electronics}).  The intrinsic time stamp
quantization floor for a single hit is $2.5/\sqrt{12} = 0.72$~ns.  This section presents a
first measurement of the timing resolution of the oWbLS pilot detector using
500~MeV proton beam data.  The analysis employs three complementary methods
: pair-difference (Method~A), per-event peak or ``proton mean time''
(Method~B), and multi-fiber averaging (Method~C), followed by per-voxel
and half-track timing studies that combine multiple fiber measurements per
reconstructed 3D position.

\subsection{Timing event selection}
\label{sec:timing_selection}

The timing analysis uses 500~MeV proton beam data from the NSRL run.  Events
are selected with a dedicated set of track-quality criteria denoted ``L4,''
the tightest selection level that retains sufficient statistics.

The 500~MeV protons are fully penetrating: at this kinetic energy the
range in water-equivalent material is $\sim$117~cm, far exceeding the 16~cm
detector depth.  The protons traverse the full Z-extent of the detector as
minimum-ionizing-like particles ($dE/dx \approx 3.1$~MeV/cm), producing
long, straight tracks that illuminate many fibers simultaneously, ideal
for intra-event timing studies.

The L4 selection proceeds in two stages.

\textit{Stage~1: floor cuts} (applied to every event):
\begin{itemize}
\item Per-hit photoelectron threshold PE $\geq 2$, suppressing sub-threshold
noise hits and dark counts.
\item At least 5 hits above the PE threshold in the event.
\item Both the XZ view (Y-fiber readout, MPPC at $-Y$ face) and ZY view
(X-fiber readout, MPPC at $-X$ face) must contain at least one hit.
\item At least 2 Z-matched voxels, where a voxel is formed by the
coincidence of an XZ-view hit at $(x, z)$ and a ZY-view hit at $(y, z)$
sharing the same Z-layer.
\item A time-cluster filter retains only hits within $\pm$50~ns of the
PE-weighted median hit time per event.  This rejects pile-up from the
beam's RF structure and stray-cluster hits surviving the initial
time-grouping window.  The PE-weighted median is computed by sorting hits
in time, accumulating PE weights, and finding the time at which the
cumulative weight reaches 50\%.
\end{itemize}

\textit{Stage~2: L4 track-quality cuts} (applied to the PCA-reconstructed
track).  Three-dimensional voxels are reconstructed by matching XZ and ZY
fiber hits at the same Z-layer.  A PE-weighted principal component analysis (PCA) is
performed on the reconstructed voxel positions: the charge-weighted covariance
matrix $C_{ij} = \sum_k w_k \, (r^k_i - \bar{r}_i)(r^k_j - \bar{r}_j) /
\sum_k w_k$ (where $w_k$ is the voxel PE and $\bar{r}$ the charge centroid)
is diagonalized to obtain the eigenvalues $\lambda_1 \leq \lambda_2 \leq
\lambda_3$ and eigenvectors.  The L4 cuts on the PCA output are:
\begin{itemize}
\item PCA linearity $(\lambda_3 - \lambda_2)/\lambda_3 > 0.92$.  This ratio
measures how elongated the charge distribution is; values near unity select
clean single-track topologies.
\item PCA residual RMS $< 1.4$~cm.  The residual is the root-mean-square
perpendicular distance of each voxel from the PCA principal axis.  This
rejects events with significant transverse scatter or multi-track
contamination. 
\item Number of Z-layer matches $\geq 8$.  Since each Z-layer corresponds
to 1~cm along the beam direction, this ensures the reconstructed track
spans at least half the detector depth.
\end{itemize}

The L4 selection yields about 200 clean proton tracks containing a total of 13{,}409
reconstructed hits across the XZ and ZY views.  The XY view (Z-fiber readout)
is excluded from the timing analysis because the Z-fibers run parallel to the beam
direction, so a beam proton illuminates essentially one Z-fiber for the
entire 16~cm track length, providing no longitudinal timing segmentation and
contributing only a single time measurement per event.
The selection criteria and yields are summarized in
table~\ref{tab:timing_selection}.

\begin{table}[htbp]
\centering
\caption{Summary of L4 timing event selection for 500~MeV proton beam data.
The floor cuts are applied first, followed by the PCA-based track-quality
cuts.}
\label{tab:timing_selection}
\begin{tabular}{lll}
\toprule
\textbf{Stage} & \textbf{Cut} & \textbf{Value} \\
\midrule
\multirow{5}{*}{Floor}
 & Per-hit PE threshold & $\geq 2$~PE \\
 & Minimum hits per event & $\geq 5$ \\
 & View requirement & XZ and ZY both populated \\
 & Z-matched voxels & $\geq 2$ \\
 & Time-cluster window & $\pm$15~ns of PE-weighted median \\
\midrule
\multirow{3}{*}{L4 quality}
 & PCA linearity $(\lambda_3-\lambda_2)/\lambda_3$ & $> 0.92$ \\
 & PCA residual RMS & $< 1.4$~cm \\
 & Z-layer matches & $\geq 8$ \\
\bottomrule
\end{tabular}
\end{table}

\subsection{Geometry and time-walk corrections}
\label{sec:timing_corrections}

Two systematic effects must be corrected before extracting timing residuals:
the position-dependent time-of-arrival variations from the finite proton
velocity and fiber propagation delay, and the amplitude-dependent time-walk
of the CITIROC discriminator.

\textit{Geometry correction.}  The raw hit time $t_{\text{raw}}$ recorded in time stamp includes two position-dependent contributions: (i)~the proton
time-of-flight from the track centroid to the hit position, and (ii)~the
propagation time of the wavelength-shifted photon along the WLS fiber from
the scintillation point to the MPPC.  For a 500~MeV proton with kinetic
energy $T = 500$~MeV ($\gamma = 1 + T/m_p = 1.533$, $\beta = 0.758$,
$v_\text{proton} = \beta c \approx 22.7$~cm/ns; conservatively taken as
$v_\text{proton} = 21.5$~cm/ns to account for energy loss along the track),
the geometry-corrected time is
\begin{equation}
t_{\text{corr}} \;=\; t_{\text{raw}} - \frac{s}{v_{\text{proton}}}
- \frac{d_{\text{fiber}}}{v_{\text{fiber}}}\,,
\label{eq:geom_corr}
\end{equation}
where $s$ is the signed projection of the hit position onto the PCA track
direction relative to the charge centroid:
\begin{equation}
s \;=\; (\vec{r}_{\text{hit}} - \vec{r}_{\text{centroid}}) \cdot \hat{e}_3\,,
\label{eq:s_proj}
\end{equation}
with $\hat{e}_3$ being the PCA eigenvector corresponding to the largest
eigenvalue (the track direction).  The fiber propagation distance
$d_{\text{fiber}}$ is the distance from the hit position to the MPPC along
the fiber.  For view~1 (XZ, Y-fibers running along $Y$, MPPC at the $-Y$
face), $d_{\text{fiber}} = y + 0.5$~cm where $y$ is the fiber row index.  For
view~2 (ZY, X-fibers running along $X$, MPPC at the $-X$ face),
$d_{\text{fiber}} = x + 0.5$~cm.  The effective group velocity in the Y11
wavelength-shifting fiber is $v_{\text{fiber}} = 18$~cm/ns, consistent with
published values for Kuraray Y11 multi-clad fiber~\cite{SFGD_prototype}.

\textit{Time-walk correction.}  The CITIROC fast-shaper output is a unipolar
shaped pulse whose leading edge crosses the discriminator threshold at a time
that depends on the pulse amplitude.  Larger pulses (higher PE) cross the
threshold earlier than smaller pulses, introducing a PE-dependent timing bias
known as ``time walk.''  A third-order polynomial walk correction $w(v, A)$
is fitted independently for each fiber view ($v = 1, 2$) using the
geometry-corrected residuals $\delta t = t_{\text{corr}} - \mu_{\text{event}}$
as a function of the hit amplitude $A$ (in PE), restricted to the range
$2 \leq A \leq 100$~PE:
\begin{equation}
w(v, A) \;=\; c_3^{(v)} A^3 + c_2^{(v)} A^2 + c_1^{(v)} A + c_0^{(v)}\,.
\label{eq:walk_poly}
\end{equation}
The event mean time $\mu_{\text{event}}$ used for the walk calibration is
obtained from a PE-weighted Gaussian fit to the distribution of
geometry-corrected hit times within each track (see Method~B below).  Outlier
hits beyond $3\sigma$ from the mean are removed iteratively before the
polynomial fit.  The corrected time is then
\begin{equation}
t \;=\; t_{\text{corr}} - w(v, A)\,.
\label{eq:walk_corr}
\end{equation}

\textcolor{black}{The fitted walk correction for the L4 proton sample is shown in
figure~\ref{fig:timing_walk} for the ZY view (X-fibers).  The scatter plot
displays the geometry-corrected timing residual $t_{\text{hit}} - t_0$ as a
function of PE, with the 3rd-order polynomial fit overlaid.}  The walk correction is most significant at low PE ($A < 10$),
where time walk can reach several nanoseconds, and becomes negligible above
$\sim$30~PE.

\textcolor{blue}{
\begin{figure}[htbp]
\centering
\includegraphics[width=0.7\textwidth]{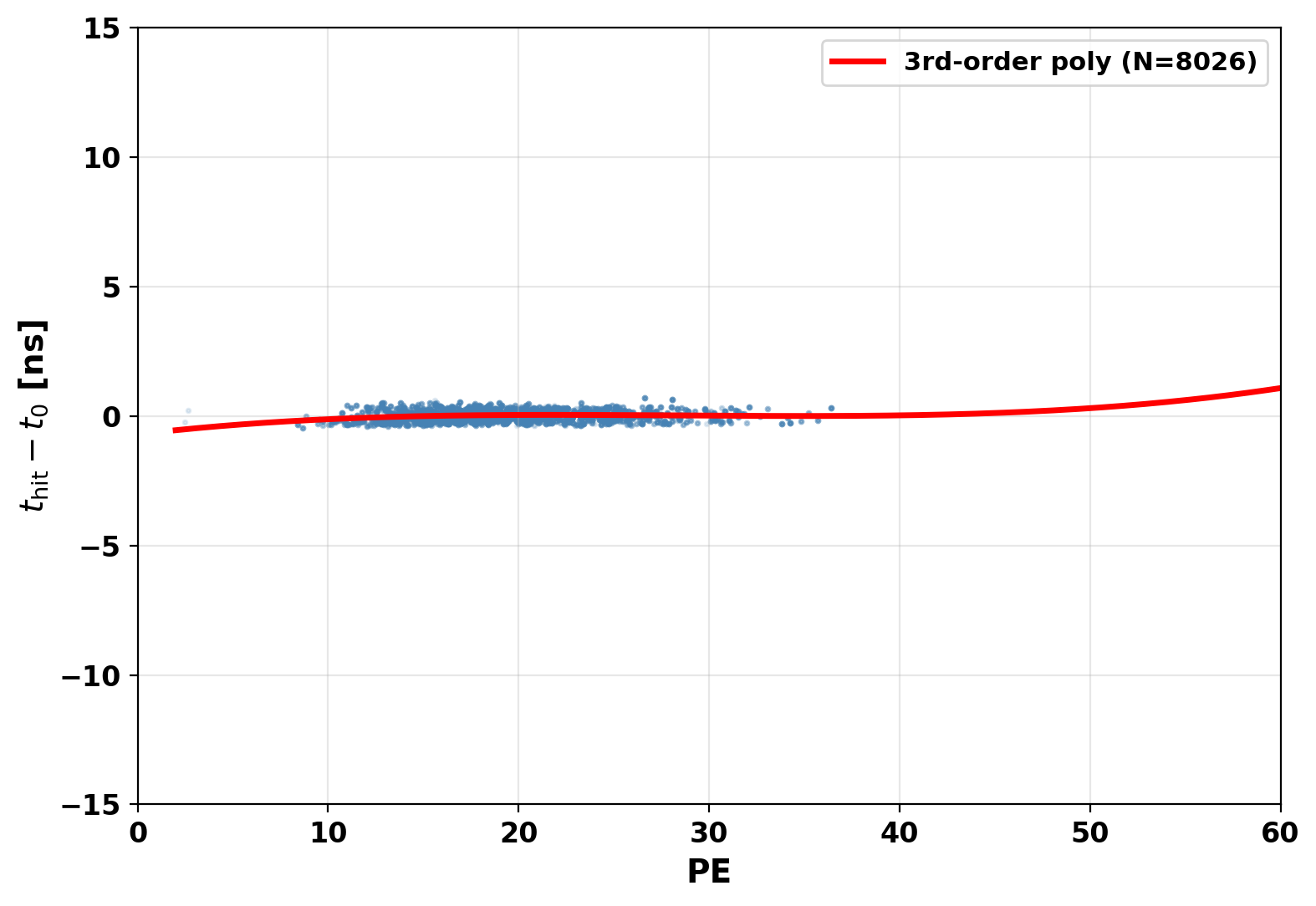}
\caption{Time-walk correction for the ZY view (X-fibers) of the L4 500~MeV
proton sample ($N = 8{,}026$ hits).  Each point represents a single hit after
geometry correction.  The red curve is the 3rd-order polynomial walk fit
(equation~\ref{eq:walk_poly}).  The walk correction is largest at low PE
($A < 10$), reaching $\sim$1~ns, and diminishes to $< 0.1$~ns above
$\sim$30~PE.}
\label{fig:timing_walk}
\end{figure}
}

\subsection{Single-channel timing: pair-difference method (Method~A)}
\label{sec:timing_pair}

The pair-difference method extracts the single-channel timing resolution
without requiring an external time reference or an event-by-event fitted
reference time.  For each track, all pairs of hits $(i, j)$ whose PE
values fall within the same PE bin are formed, and the pairwise time
difference $\Delta t_{ij} = t_i - t_j$ is computed after geometry and
walk corrections.  If both hits have the same intrinsic timing resolution
$\sigma_t$, the distribution of $\Delta t$ has width
$\sigma_{\Delta t} = \sqrt{2}\,\sigma_t$, so the single-channel resolution
is
\begin{equation}
\sigma_t \;=\; \frac{\sigma(\Delta t)}{\sqrt{2}}\,,
\label{eq:pair_diff}
\end{equation}
where $\sigma(\Delta t)$ is the clipped standard deviation of the
pair-difference distribution.  The clipping procedure performs two iterations
of $3\sigma$ outlier rejection: at each iteration, the mean and standard
deviation are computed, and values beyond $3\sigma$ from the mean are
removed.  This suppresses the tails from residual pile-up or noise while
retaining the core of the distribution.

Table~\ref{tab:timing_pair} shows the pair-difference results for three
PE bins containing sufficient statistics (at least 10 pairs).  The bins at
$[2, 4)$ PE (38 pairs, $\sigma_t = 3.6$~ns) and $[6, 8)$ PE (61 pairs,
$\sigma_t = 0.01$~ns) are excluded from the table due to limited statistics.
The dominant PE bin in the 500~MeV proton sample is $[12, 20)$~PE,
corresponding to the most probable energy loss per fiber for a
minimum-ionizing proton at this energy.  With 372{,}882 pairs, the
single-channel resolution in this bin is $\sigma_t = 0.17$~ns.  The
resolution improves monotonically with PE, reaching $\sigma_t = 0.14$~ns in
the $[20, 40)$~PE bin.

\begin{table}[htbp]
\centering
\caption{Method~A pair-difference timing results for 500~MeV proton beam data
(L4 selection).  $\sigma_{\text{pair}}$ is the clipped standard
deviation ($3\sigma$, 2 iterations) of the pairwise time differences;
$\sigma_t = \sigma_{\text{pair}}/\sqrt{2}$ is the inferred single-channel
resolution.}
\label{tab:timing_pair}
\begin{tabular}{crrr}
\toprule
\textbf{PE range} & \textbf{$N_{\text{pairs}}$} &
\textbf{$\sigma_{\text{pair}}$ [ns]} & \textbf{$\sigma_t$ [ns]} \\
\midrule
$[8, 12)$   &  25{,}115  & 0.27 & 0.19 \\
$[12, 20)$  & 372{,}882  & 0.24 & 0.17 \\
$[20, 40)$  &  82{,}348  & 0.20 & 0.14 \\
\bottomrule
\end{tabular}
\end{table}

\subsection{Single-channel timing: proton mean time method (Method~B)}
\label{sec:timing_peak}

Method~B defines a per-event reference time from the track itself.  For each
of the L4 proton tracks, the geometry-corrected hit times are histogrammed
with PE weights and fitted with a Gaussian.  The fitted mean
$\mu_{\text{event}}$ serves as the ``proton mean time'' --- the
PE-weighted average arrival time of all photons associated with the track.
For events with fewer than 6 hits, the PE-weighted median is used instead of
a Gaussian fit.

The timing residual for each hit is then
\begin{equation}
\delta t \;=\; t_{\text{hit}} - \mu_{\text{event}}\,,
\label{eq:residual}
\end{equation}
where $t_{\text{hit}}$ is the walk-corrected and geometry-corrected time.
The residuals from all hits across all tracks are pooled into a
single distribution and fitted with a Gaussian to obtain the overall
single-channel resolution.

The pooled residual distributions are shown in
figure~\ref{fig:timing_residual}.  The Gaussian fit to the full distribution
(left panel) yields $\sigma = 0.20$~ns, while the distribution restricted to
hits with PE~$> 10$ (right panel, 13{,}243 hits) yields $\sigma = 0.24$~ns.
The slightly larger width for the PE~$> 10$ subset reflects the narrower
PE range: the PE-weighted event mean is dominated by these high-PE hits, so
the residuals no longer benefit from the averaging effect that suppresses the
low-PE tails.  Both distributions are narrow and symmetric about zero,
confirming excellent inter-channel timing coherence.

\begin{figure}[htbp]
\centering
\includegraphics[width=0.95\textwidth]{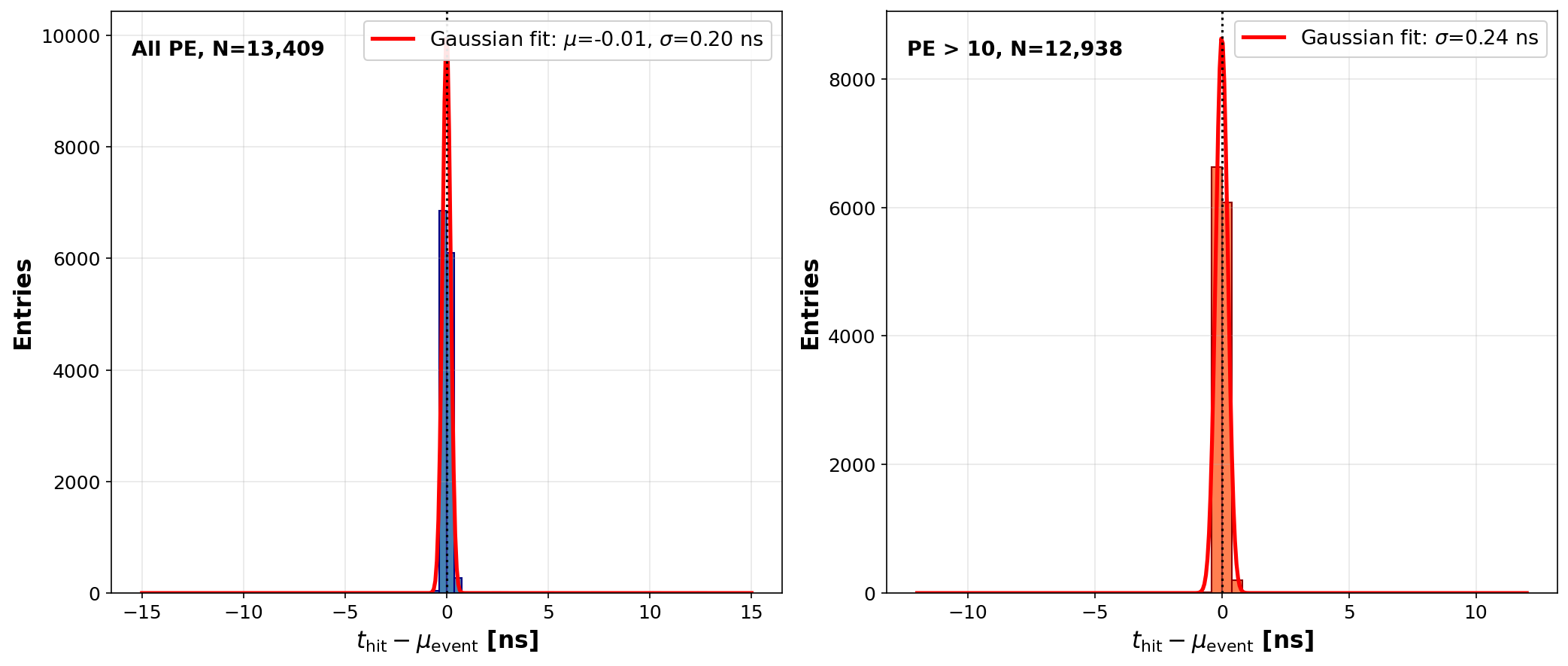}
\caption{Pooled timing residual distributions $\delta t = t_{\text{hit}} -
\mu_{\text{event}}$ for 500~MeV proton beam data (L4 selection).
Left: all PE values ($N = 13{,}409$, Gaussian fit $\sigma = 0.20$~ns).
Right: hits with PE~$> 10$ ($N = 13{,}243$, Gaussian fit
$\sigma = 0.24$~ns).  $\mu_{\text{event}}$ is the PE-weighted Gaussian peak
time per track.}
\label{fig:timing_residual}
\end{figure}

The single-channel resolution as a function of PE is obtained by binning the
Method~B residuals into PE ranges and fitting the core of each bin's residual
distribution with a Gaussian.  Figure~\ref{fig:timing_sigma_pe} shows
$\sigma_t$ vs.\ PE for Method~B (per-event peak, blue circles).
The resolution improves steeply
from $\sigma_t \approx 8$~ns at $\sim$3~PE (near the noise threshold) to
$\sigma_t \approx 0.28$~ns at $\sim$30~PE.

The PE dependence of the Method~B resolution is described by a power-law
fit
\begin{equation}
\sigma_t(A) \;=\; \frac{q_0}{A^{q_1}} + q_2\,,
\label{eq:sigma_pe_fit}
\end{equation}
with best-fit parameters $q_0 = 21.1 \pm 17.8$~ns, $q_1 = 1.27 \pm 0.40$,
and $q_2 \approx 0$~ns (consistent with zero within uncertainties).  The
power-law exponent $q_1 \approx 1.3$ is steeper than the $1/\sqrt{A}$
scaling expected from pure photostatistics ($q_1 = 0.5$), indicating that
the CITIROC discriminator jitter contributes significantly at low PE and
diminishes rapidly as the signal-to-noise ratio improves.  The asymptotic
floor $q_2 \approx 0$ suggests that the intrinsic electronics jitter is
well below the time stamp step at high PE.


\begin{figure}[htbp]
\centering
\includegraphics[width=0.65\textwidth]{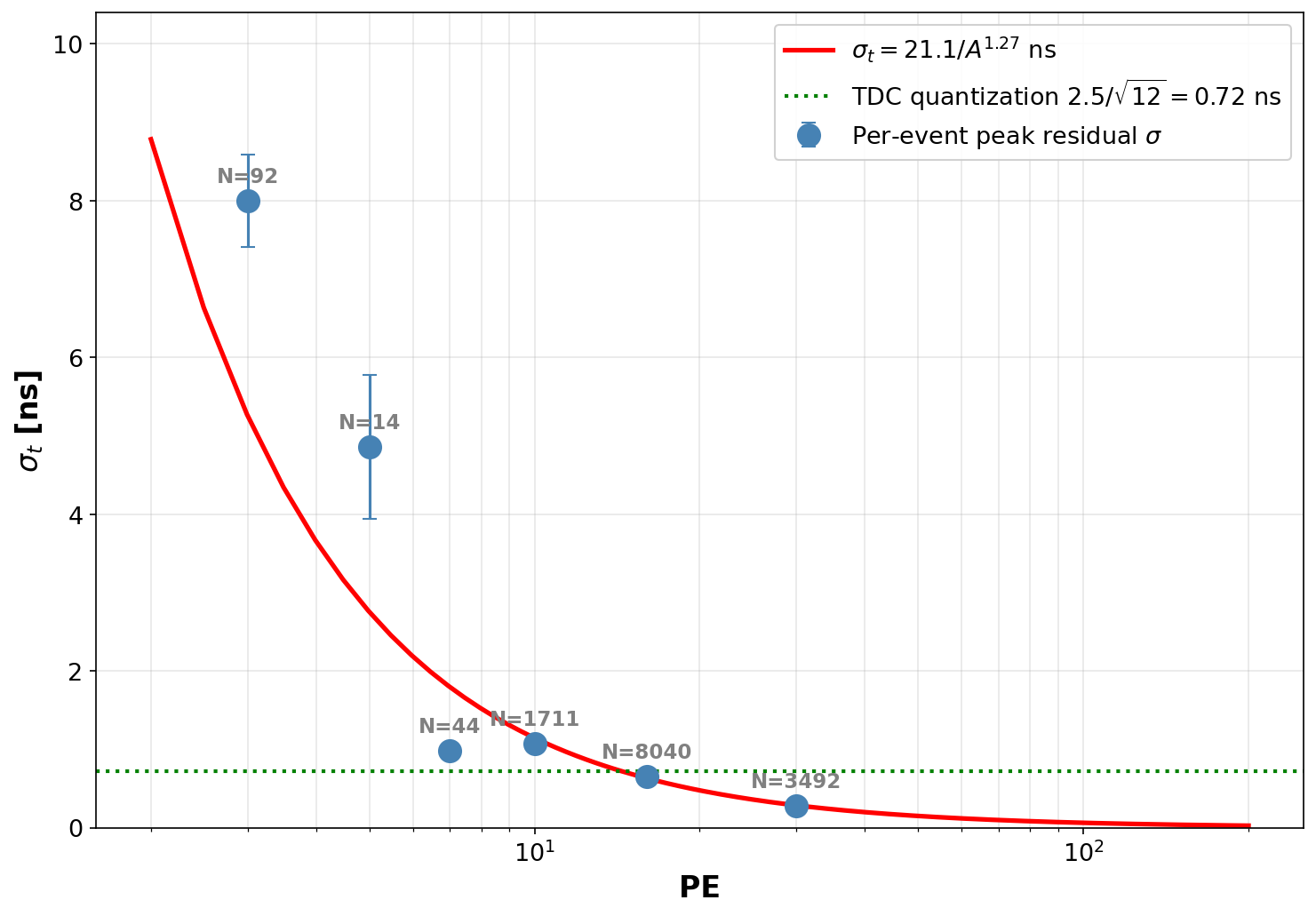}
\caption{\textcolor{black}{Single-channel timing resolution $\sigma_t$ as a
function of PE for 500~MeV proton beam data (L4 selection,
13{,}409 hits).  Blue circles: per-event peak residual $\sigma$ per PE bin
(Method~B).  Red curve: power-law fit $\sigma_t = 21.1/A^{1.27}$~ns
(equation~\ref{eq:sigma_pe_fit}).  Green dotted line: TDC quantization floor
$2.5/\sqrt{12} = 0.72$~ns.  Numbers above each point indicate the number of
hits in that PE bin.}}
\label{fig:timing_sigma_pe}
\end{figure}

\subsection{Multi-fiber averaging (Method~C)}
\label{sec:timing_multi}

Method~C tests whether the timing resolution scales as $1/\sqrt{N}$ when
averaging $N$ independent fiber measurements per reconstructed position.
This is a direct probe of whether the fiber-to-fiber timing variations are
dominated by uncorrelated jitter rather than systematic effects.

After applying geometry and walk corrections, the times of all hits in the
same fiber bar defined by the same transverse position and Z-layer in a
given view, are averaged with PE weights.  For view~1 (XZ), a bar is
identified by coordinates $(x, z)$; for view~2 (ZY), by $(y, z)$.  The
PE-weighted bar time is
\begin{equation}
\bar{t}_\text{bar} \;=\; \frac{\sum_k A_k \, t_k}{\sum_k A_k}\,,
\label{eq:bar_time}
\end{equation}
where the sum runs over all hits in the bar and $A_k$, $t_k$ are the PE
and corrected time of each hit.

\textit{1-view resolution.}  Pair differences between bar-averaged times
within the same view yield the single-view (1-fiber) timing resolution:
$\sigma_{\text{1-view}}^{(\text{XZ})} = 0.28$~ns for the XZ view
(Y-fibers) and $\sigma_{\text{1-view}}^{(\text{ZY})} = 0.19$~ns for the
ZY view (X-fibers).  The difference between views reflects the different
fiber lengths (and hence attenuation and dispersion) for the two
orientations.  The average single-view resolution is
$\sigma_{\text{1-view}} = 0.23$~ns.

\textit{2-view (per-voxel) resolution.}  The per-voxel time is formed by
averaging the bar times from the XZ and ZY views at the same $(x, y, z)$
voxel position:
\begin{equation}
t_\text{voxel} \;=\; \frac{A_\text{XZ}\,\bar{t}_\text{XZ}
  + A_\text{ZY}\,\bar{t}_\text{ZY}}{A_\text{XZ} + A_\text{ZY}}\,,
\label{eq:voxel_time}
\end{equation}
where $A_\text{XZ}$ and $A_\text{ZY}$ are the total PE in the
corresponding bars.  Pair differences between 2-view voxel times yield
$\sigma_{\text{2-view}} = 0.19$~ns.

The expected 2-view resolution from $1/\sqrt{N}$ scaling is
$\sigma_{\text{1-view}}/\sqrt{2} = 0.23/\sqrt{2} = 0.17$~ns.  The
measured 0.19~ns is consistent with this expectation within the statistical
precision, confirming that the two views provide largely independent timing
information.  The measured 0.19~ns is consistent with the expected
$\sigma_{\text{1-view}}/\sqrt{2} = 0.17$~ns, confirming that the two views
provide largely independent timing information.

\subsection{Per-voxel and half-track timing}
\label{sec:timing_voxel}

The multi-fiber averaging of Method~C establishes that combining two fiber
views per voxel improves the resolution.  This subsection extends the
analysis to the per-voxel pair-difference resolution as a function of
total voxel PE, and to the half-track timing resolution obtained by
splitting each track into two halves.

\textit{Per-voxel pair-difference.}  For each voxel, the voxel time
$t_\text{voxel}$ (equation~\ref{eq:voxel_time}) and the total voxel PE
$A_\text{voxel} = A_\text{XZ} + A_\text{ZY}$ are computed.  All pairs of
voxels within the same track and the same voxel-PE bin are formed, and the
pair-difference method (equation~\ref{eq:pair_diff}) is applied.  The L4
tracks yield 1{,}279 reconstructed voxels with timing information.  The
median voxel PE is 177, with a 90th percentile of 282 and a maximum of 563,
reflecting the summed energy deposition from a 500~MeV proton across two
fiber views at each Z-layer.
The dominant voxel-PE bin is $[80, 400)$~PE, containing 8{,}192 voxel pairs.
The per-voxel timing resolution at this PE is $\sigma_t = 0.16$~ns.  A second
bin at $[40, 80)$~PE (45 pairs) yields $\sigma_t = 0.49$~ns, but with limited
statistics.

\textit{Half-track timing.}  Each proton track is split at the median
projection along the PCA direction into two halves (``first'' and ``second''
along the track).  Only tracks with at least 4 voxels are used, and each
half must contain at least 2 voxels.  The PE-weighted mean voxel time is
computed for each half:
\begin{equation}
t_\text{half} \;=\;
\frac{\sum_{k \in \text{half}} A_k^\text{voxel} \, t_k^\text{voxel}}
     {\sum_{k \in \text{half}} A_k^\text{voxel}}\,.
\label{eq:half_time}
\end{equation}
The pair difference $\Delta t = t_\text{first} - t_\text{second}$ is formed
for each track, and the half-track timing resolution is
$\sigma_\text{half} = \sigma(\Delta t) / \sqrt{2}$, where $\sigma(\Delta t)$
is the clipped standard deviation across the L4 tracks.

For tracks with minimum half-PE $\geq 500$ (all L4 tracks qualify, with
mean half-PE = 1{,}143), the measured half-track timing resolution is
$\sigma_\text{half} = 0.05$~ns.  This represents the precision with which
two independently measured halves of the same track agree on their mean
arrival time, and demonstrates that the oWbLS detector can timestamp track
segments at the sub-100~ps level when sufficient photoelectron statistics
are available.

\subsection{Discussion of timing results}
\label{sec:timing_discussion}

All three single-channel methods yield timing resolutions well below the
single hit time stamp step floor of $2.5/\sqrt{12} = 0.72$~ns.  This is possible
because the 500~MeV proton beam produces highly synchronous energy
depositions: all fiber hits within a single track arrive within a sub-ns
window, and most channels register on the same time tick.  The intra-event
timing spread therefore probes the correlated sub-tick jitter in the CITIROC
discriminator and FPGA digitization chain, rather than the absolute time stamp
granularity.

The consistency between the three methods provides confidence in the result.
Method~A (pair-difference) is self-calibrating and independent of the event
reference time.  Method~B (proton mean time) depends on the quality of the
per-event Gaussian fit but provides finer PE binning.  Method~C
(multi-fiber averaging) tests the statistical independence of fiber-to-fiber
measurements and confirms the expected $1/\sqrt{N}$ improvement.  The
agreement between all three methods at $\sigma_t \approx 0.2$~ns
(depending on PE range) demonstrates that the result is robust against
analysis choices.

The per-voxel resolution of 0.16~ns at PE~$\sim$177 and the half-track
resolution of 0.05~ns at PE~$\sim$1{,}143 demonstrate the benefit of
aggregating photoelectron statistics: as more fibers contribute to a
reconstructed position or track segment, the timing precision improves
dramatically.  The half-track precision of 0.05~ns corresponds to a
time-of-flight measurement precision of $\Delta L = v_\text{proton} \times
\sigma_\text{half} \approx 22.7 \times 0.05 \approx 1.1$~cm along the
track, comparable to the 1~cm fiber pitch.
The key timing results are summarized in table~\ref{tab:timing_summary}.

\begin{table}[htbp]
\centering
\caption{Summary of timing resolution measurements for 500~MeV proton beam
data (L4 selection).  Single-channel methods
measure individual fiber timing precision; multi-fiber and per-voxel
methods combine 2 fiber views; half-track aggregates over $\sim$7 voxels
per half.}
\label{tab:timing_summary}
\begin{tabular}{llrl}
\toprule
\textbf{Method} & \textbf{PE range} & \textbf{$N$} &
\textbf{$\sigma_t$ [ns]} \\
\midrule
\multicolumn{4}{l}{\textit{Single-channel: pair-difference (Method~A)}} \\
 & $[8, 12)$~PE  & 25{,}115 pairs & 0.19 \\
 & $[12, 20)$~PE & 372{,}882 pairs & 0.17 \\
 & $[20, 40)$~PE & 82{,}348 pairs & 0.14 \\
\midrule
\multicolumn{4}{l}{\textit{Single-channel: proton mean time (Method~B)}} \\
 & Pooled, all PE & 13{,}409 hits & 0.20 \\
 & Pooled, PE $> 10$ & 13{,}243 hits & 0.24 \\
 & Power-law fit & & $21.1/A^{1.27}$ \\
\midrule
\multicolumn{4}{l}{\textit{Multi-fiber averaging (Method~C)}} \\
 & 1-view (XZ, Y-fibers) & & 0.28 \\
 & 1-view (ZY, X-fibers) & & 0.19 \\
 & 1-view average & & 0.23 \\
 & 2-view (per voxel) & & 0.19 \\
 & Expected ($\sigma_1/\sqrt{2}$) & & 0.17 \\
\midrule
\multicolumn{4}{l}{\textit{Per-voxel pair-difference}} \\
 & Voxel PE $\in [80, 400)$ & 8{,}192 pairs & 0.16 \\
\midrule
\multicolumn{4}{l}{\textit{Half-track}} \\
 & Half PE $\geq 500$ & all L4 tracks & 0.05 \\
\bottomrule
\end{tabular}
\end{table}

}

\section{Summary and outlook}
\label{sec:conclusion}

We have described the design, construction, and beam test of a
pilot 3D-projection detector based on opaque water-based liquid scintillator.
The detector, with an active volume of $8 \times 8 \times 16$~cm$^3$, was
constructed at BNL using acrylic boards bonded with Weld-On solvent cement,
with fiber penetrations sealed using Weld-On 40 reactive acrylic adhesive.
The detector was instrumented with 320 Kuraray Y11 multi-clad WLS fibers in
three orthogonal views, and read out by Hamamatsu S13360-1325CS MPPCs coupled
to CITIROC-based front-end electronics inherited from the Baby~MIND and
SuperFGD detectors of the T2K experiment.

The detector was commissioned with cosmic ray data at BNL and subsequently
exposed to 50, 100, 250, and 500~MeV proton beams at the NASA Space Radiation
Laboratory.  The 50 and 100~MeV data sets have limited statistics; the 250 and
500~MeV data sets provide the primary physics results.  The MPPC gain
calibration was performed using a pulsed LED system in 2020, with per-channel
gain constants applied throughout the analysis; the
systematic uncertainty on the absolute PE scale from the five-year gap between
calibration and beam test is estimated at $\pm$2--3\%, dominated by
unmonitored temperature differences.  Cosmic muon data confirmed a light yield
of approximately 13--14~PE per MIP traversal per fiber.

Three-dimensional event reconstruction using the three-view fiber projection
technique was demonstrated for both cosmic muon tracks and proton beam events.
The 50~MeV protons stop within approximately 2.2~cm, the
100~MeV protons stop inside the detector (range $\sim$7.7~cm), while the
250 and 500~MeV protons are fully penetrating, with ranges far exceeding the
16~cm detector depth.
A detailed study of the transverse light spread was performed by comparing
radial charge profiles in data with Geant4 Monte Carlo predictions.
\textcolor{black}{The measured profiles are sharply
peaked, with $\sim$94\% of the charge contained within 1~cm of the beam axis.
A direct comparison with a simulation using a 2~cm scattering length shows
that the data exhibit significantly tighter confinement, placing the
effective scattering length of the oWbLS medium well below 2~cm.}  Both the
250 and 500~MeV data sets show the same degree of
confinement, despite their $\sim$50\% difference in $dE/dx$, confirming that
the measurement reflects intrinsic optical properties of the medium.

\textcolor{black}{A first measurement of the timing resolution was performed
using 500~MeV proton beam data with a tight track-quality selection (L4).
Three complementary methods: pair-difference, per-event peak (proton mean
time), and multi-fiber averaging, yield consistent single-channel timing
resolutions in the range $\sigma_t \approx 0.14$--$0.28$~ns depending on PE,
per-voxel resolution of $\sim$0.16~ns, and half-track resolution of
$\sim$0.05~ns.}
The key parameters and measured performance are
summarized in table~\ref{tab:summary}.
\begin{table}[htbp]
\centering
\caption{Summary of the pilot oWbLS detector parameters and measured
performance.}
\label{tab:summary}
\begin{tabular}{ll}
\toprule
\textbf{Parameter} & \textbf{Value} \\
\midrule
Active volume & $8 \times 8 \times 16$~cm$^3$ \\
Active medium & Opaque WbLS (oWbLS) \\
Scintillation light yield (oWbLS) & \textcolor{black}{$\sim$12,000~photons/MeV} \\
Fiber type & Kuraray Y11(200), multi-clad, 1.0~mm diameter \\
Fiber pitch & 10~mm \\
Number of fiber views & 3 (X, Y, Z) \\
Total fibers & 320 \\
Photosensor & Hamamatsu S13360-1325CS MPPC \\
MPPC active area & $1.3 \times 1.3$~mm$^2$ \\
MPPC pixel pitch & 25~$\mu$m (2,668 pixels) \\
Readout electronics & CITIROC-based FEB (Baby~MIND/SuperFGD) \\
Timing resolution & 2.5~ns \\
Light yield (cosmic MIP) & $\sim$13--14~PE/fiber \\
Effective voxel size & $1 \times 1 \times 1$~cm$^3$ \\
Beam test energies & 50, 100, 250, 500~MeV protons \\
 & (50, 100~MeV limited statistics) \\
Scattering length & \textcolor{black}{$< 2$~cm (data more confined than 2~cm MC)} \\
\textcolor{black}{Timing $\sigma_t$ (single channel, 12--20~PE)} &
  \textcolor{black}{0.17~ns (pair-diff)} \\
\textcolor{black}{Timing $\sigma_t$ (per voxel, PE $\geq 80$)} &
  \textcolor{black}{0.16~ns} \\
\textcolor{black}{Timing $\sigma_t$ (half-track, PE $\geq 500$)} &
  \textcolor{black}{0.05~ns} \\
Gain calibration systematic & $\pm$2--3\% \\
\bottomrule
\end{tabular}
\end{table}
These results demonstrate the viability of the 3D-projection oWbLS detector
concept as a scalable alternative to mechanically segmented fine-grained
scintillator detectors.  Future work will focus on constructing a series of
modular prototype detectors, each approximately 20~cm on a side, with fiber
pitches of 1--2~cm and several hundred readout channels per module.  If funding
permits through the DOE LAB-26-3575 program, up to six such modules will be
built and tested, with a total of approximately 4,500 readout channels.  These
modules will be filled with heavy-metal-loaded oWbLS formulations (Pb and W at
loadings of 5--30~wt.\%) to tune the radiation length for electromagnetic and
hadronic calorimetry studies.  Extended beam tests are planned at NSRL with
proton energies spanning 50--2500~MeV, as well as at the BNL Accelerator Test
Facility with 75~MeV electron beams.  The technology is being evaluated for
potential applications in neutrino near detectors, rare-process searches, and
collider calorimetry.

\section*{Acknowledgments}

This work was supported by the U.S. Department of Energy under Contract No. DE-SC0012704.
This work was also supported by the Laboratory Directed Research and Development
(LDRD) program of Brookhaven National Laboratory.  The beam test was performed
at the NASA Space Radiation Laboratory at BNL.  We thank the NSRL operations
staff Xiaodong Jiang and Mike Sivertz for their support during the beam test campaigns.  The readout
electronics were provided through the US--Japan cooperative program in high
energy physics.  We thank the Baby~MIND and SuperFGD collaborations for the
development of the CITIROC-based front-end board system.


\end{document}